%% file: main.tex
 \documentclass[english]{jssst_ppl} %% English

\title{Lexicographic Ranking Supermartingales with \\ Lazy Lower Bounds}
\author{Toru Takisaka$^1$, Libo Zhang$^2$, Changjiang Wang$^1$, Jiamou Liu$^2$}
\inst{%
$^1$ University of Electronic Science and Technology of China\\
\texttt{takisaka@uestc.edu.cn; 202222080938@std.uestc.edu.cn}
\medskip\par%
$^2$ The University of Auckland\\
\texttt{lzha797@aucklanduni.ac.nz; jiamou.liu@auckland.ac.nz}
}

\input{userPreambles}

\begin{document}
\maketitle

\input{articleBody}

\end{document}

%% file: userPreambles.tex
\usepackage{amssymb} %disabled upon POPL transition
\usepackage{amsmath}
\usepackage{graphicx}
\usepackage[ruled,linesnumbered,noend]{algorithm2e}
\usepackage{comment}
     \excludecomment{cav} % activate this to show the entire body
     \includecomment{arxiv} % use this contents for arXiv version
    \excludecomment{full} % use to hide contents for CAV conference submission  
    \excludecomment{ver0204}
    \excludecomment{blue}
\usepackage{tabu}
\usepackage{multirow}
\usepackage{float}
\usepackage{wrapfig}
\usepackage{longtable}
\usepackage{paralist}
\usepackage{stmaryrd}
\usepackage{listings}
\usepackage[disable,colorinlistoftodos,textsize=footnotesize]{todonotes}
\usepackage{fontawesome}

\usepackage{tikz}
\usetikzlibrary{automata, positioning, arrows, shapes.geometric}
\tikzset{
->, % makes the edges directed
>=stealth', % makes the arrow heads bold
node distance=2cm, % specifies the minimum distance between two nodes. Change if necessary.
every state/.style={thick}, % sets the properties for each ¡Çstate¡Ç node
initial text=$ $, % sets the text that appears on the start arrow
}

\usepackage{apxproof}
\theoremstyle{plain}
\newtheoremrep{mytheorem}{Theorem}[section]
\newtheoremrep{myproposition}[mytheorem]{Proposition}
\newtheoremrep{mylemma}[mytheorem]{Lemma}
\newtheoremrep{mycorollary}[mytheorem]{Corollary}
\theoremstyle{definition}
\newtheoremrep{myexample}[mytheorem]{Example}
\newtheoremrep{myremark}[mytheorem]{Remark}
\newtheoremrep{mydefinition}[mytheorem]{Definition}

\usepackage{array}
\newcolumntype{C}[1]{>{\centering\let\newline\\\arraybackslash\hspace{0pt}}m{#1}}
\newcolumntype{R}[1]{>{\raggedleft\let\newline\\\arraybackslash\hspace{0pt}}m{#1}}
\newcolumntype{L}[1]{>{\raggedright\let\newline\\\arraybackslash\hspace{0pt}}m{#1}}

\newcommand{\Av}{\mathrm{A\kern-0.09em v}}

\newcommand{\Lv}{\mathsf{Lv}}
\newcommand{\mysucc}{\mathrm{succ}}

\newcommand{\myup}{\mathit{Up}}
\newcommand{\sem}[1]{\llbracket #1 \rrbracket}
\newcommand{\br}[1]{\{ #1 \}}
\newcommand{\seqOf}[1]{(#1)_{t=0}^\infty}

\newcommand{\mychar}{{\bf 1}}

\newcommand{\bfX}{\mathbf{X}}

\newcommand{\bbE}{\mathbb{E}}

\newcommand{\bbN}{\mathbb{N}}

\newcommand{\bbP}{\mathbb{P}}

\newcommand{\bbR}{\mathbb{R}}

\newcommand{\bbX}{\mathbb{X}}

\newcommand{\calB}{\mathcal{B}}
\newcommand{\calC}{\mathcal{C}}
\newcommand{\calD}{\mathcal{D}}

\newcommand{\calF}{\mathcal{F}}

\newcommand{\calI}{\mathcal{I}}

\newcommand{\calS}{\mathcal{S}}
\newcommand{\calT}{\mathcal{T}}

\usepackage{subfig}
\usepackage{stfloats}%for correct floating of two-columns figures
\usepackage{footmisc}

\usepackage[all,pdftex]{xy}
\SelectTips{cm}{}
\xyoption{rotate}

\usepackage{url}
\usepackage[colorlinks=true,driverfallback=dvipdfm]{hyperref}

%% file: articleBody.tex
\begin{abstract}
\emph{Lexicographic Ranking SuperMartingale}  (LexRSM) is a probabilistic extension of \emph{Lexicographic Ranking Function}  (LexRF), which is a widely accepted technique for verifying program termination. In this paper, we are the first to propose sound probabilistic extensions of LexRF with a weaker non-negativity condition, called \emph{single-component} (SC) non-negativity. It is known that such an extension, if it exists, will be nontrivial due to the intricacies of the probabilistic circumstances.

Toward the goal,
we first devise the notion of \emph{fixability}, which offers a systematic approach for analyzing the soundness of possibly negative LexRSM. This notion yields a desired extension of LexRF that is sound for general stochastic processes. We next propose another extension, called \emph{Lazy LexRSM}, toward the application to automated verification; it is sound over probabilistic programs with linear arithmetics, while its subclass is amenable to automated synthesis via linear programming.
We finally propose a LexRSM synthesis algorithm for this subclass, and perform experiments.
\end{abstract}

\section{Introduction}\label{sect:introduction}
{\bf Background 1: Lexicographic RFs with different non-negativity conditions.} 
\emph{Ranking function} (RF) is one of the most well-studied tools for verifying program termination. 
An RF is typically a real-valued function over program states that satisfies: 
\begin{inparaenum}[(a)]
\item the \emph{ranking condition}, which requires 
an RF to decrease its value by a constant through each transition; and \item the \emph{non-negativity condition}, which imposes a
lower bound on the value of the RF so that its infinite descent through transitions is prohibited.
\end{inparaenum}
The existence of such a function implies termination of the underlying program, and therefore, 
one can automate verification of program termination by RF synthesis algorithms.

Improving the \emph{applicability} of RF synthesis algorithms, i.e., making them able to prove termination of a wider variety of programs, 
is one of the core interests in the study of RF.  
A \emph{lexicographic extension} of RF (LexRF)~\cite{BradleyMS05,Ben-AmramG15} is known as a simple but effective approach 
to the problem. 
Here, a LexRF is a function to real-valued \emph{vectors} instead of the reals, and its ranking condition is imposed with respect to the lexicographic order. 
For example, the value of a LexRF 
may change from 
$(1,1,1)$ to $(1,0,2)$ 
through a state transition; 
here, the value 
``lexicographically decreases by 1'' through the transition,  
that is, it decreases by 1 in some dimension while it is non-increasing on the left to that dimension.  
LexRF is particularly good at handling nested structures of programs, as vectors can measure the progress of different ``phases'' of programs separately.
LexRF is also used in top-performing termination provers (e.g.,~\cite{ultimateAutomizer}).

There are several known ways to impose non-negativity on LexRFs (see also Fig.~\ref{fig:ExForGLexRSM}):
\begin{inparaenum}[(a)]
    \item     \emph{Strong non-negativity}, 
    which requires non-negativity in every dimension of the LexRF; 
        \item         \emph{leftward non-negativity}, 
        which requires non-negativity on the left of the 
                \emph{ranking dimension} of each transition, i.e., the dimension where the value of the LexRF should strictly decrease through the transition; and
    \item      \emph{single-component non-negativity}, 
    which requires non-negativity only in the ranking dimensions.
\end{inparaenum}
It is known that any of these non-negativity conditions makes the resulting LexRF \emph{sound}~\cite{BradleyMS05,Ben-AmramG15}, i.e., a program indeed terminates whenever it admits a LexRF with either of these non-negativity conditions. For better applicability, single-component non-negativity is the most preferred, as it is the weakest constraint among the three.

\input{figures/ExForGLexRSM.tex}

\smallskip 
\noindent
{\bf Background 2: Probabilistic programs and lexicographic RSMs.} 
One can naturally think of a probabilistic counterpart of the above argument. 
One can consider \emph{probabilistic programs} that admit randomization
in conditional branching and variable updates. 
The notion of RF is then generalized to \emph{Ranking SuperMartingale} (RSM), 
a function similar to RFs except that
the ranking condition requires an RSM to decrease its value \emph{in expectation}. 
The existence of an RSM typically implies \emph{almost-sure termination} of the underlying program, i.e., termination of the program with probability 1.

Such a probabilistic extension has been actively studied, in fact: 
probabilistic programs are used in e.g.,  stochastic network protocols~\cite{parker2013verification}, randomized algorithms~\cite{dubhashi2009concentration,karp1991introduction},
security~\cite{barthe2016proving,lobo2021programming,barthe2016programming}, and planning~\cite{canal2019probabilistic}; and there is a rich body of studies 
in RSM as a tool for automated verification of probabilistic programs (see \S\ref{sect:relWork}). 
Similar to the RF case, 
a lexicographic extension of RSM (\emph{LexRSM},~\cite{AgrawalCP18,ChatterjeeGNZZ21}) is an effective approach to improve its applicability.
In addition to its advantages over nested structures, LexRSM can also witness almost-sure termination of certain probabilistic programs with infinite expected runtime~\cite[Fig. 2]{AgrawalCP18}; certifying such programs is known as a major challenge for RSMs.

\medskip
\noindent

{\bf Problem: Sound probabilistic extension of LexRF with weaker non-negativity.} 
strongly non-negative LexRF soundly extends to LexRSM in a canonical way~\cite{AgrawalCP18}, i.e., 
basically by changing the clause ``decrease by a constant'' in the ranking condition of LexRF to ``decrease by a constant \emph{in expectation}''.  
In contrast, 
the similar extension of leftward or single-component non-negative LexRF yields an \emph{unsound} LexRSM notion~\cite{ferrer2015probabilistic, ChatterjeeGNZZ21}. 
To date, a sound LexRSM with the weakest 
non-negativity in the literature is \emph{Generalized LexRSM} (GLexRSM)~\cite{ChatterjeeGNZZ21},
which demands leftward non-negativity \emph{and} an additional one, so-called \emph{expected leftward non-negativity}. 
Roughly speaking, the latter requires LexRSMs to be non-negative in each dimension  (in expectation) upon ``exiting'' the left of the ranking dimension. For example, in Fig.~\ref{fig:ExForGLexRSM}, it requires $b_2$ to be non-negative, as the second dimension of $\boldsymbol{\eta}$ ``exits'' the left of the ranking dimension upon the transition $\ell_1 \rightarrow \ell_2$. 
GLexRSM does not generalize either leftward or single-component non-negative LexRF, in the sense that the former is strictly more restrictive than the latter two when it is considered over non-probabilistic programs.

These results do not mean that leftward or single-component non-negative LexRF can never be extended to LexRSM, however. More concretely, the following problem is valid 
(see the last paragraph of $\S$\ref{sect:preliminaries} for a formal argument):

\medskip

\fbox{\parbox[c]{0.9\linewidth}{
    \emph{KEY PROBLEM: Find a sound LexRSM notion that} instantiates\footnotemark \emph{single-component non-negative LexRF, i.e., a LexRSM notion 
        whose condition is no stronger than that of single-component non-negative LexRF in
    non-probabilistic settings.}
}}\footnotetext{
    We use the term 
    ``instantiate'' to emphasize that we compare LexRSM and LexRF.         } 

\medskip
\noindent
We are motivated to study this problem for a couple of reasons.
First, it 
is a paraphrase of the following fundamental question: 
\emph{when do negative values of (Lex)RSM cause trouble}, say, to its soundness? 
This question is a typical example in the study of RSM where the question becomes challenging due to its probabilistic nature. The question also appears in other topics in RSM; for example, 
it is known that the classical variant rule of Floyd-Hoare logic does not extend to almost-sure termination of probabilistic programs in a canonical way~\cite{Huang0CG19}, due to the complicated treatment of negativity in RSMs.
To our knowledge, this question has only been considered in an ad-hoc manner through counterexamples (e.g.,~\cite{ferrer2015probabilistic,Huang0CG19,ChatterjeeGNZZ21}), and we do not yet have a systematic approach to answering it. 
\input{figures/LeadingEx}

Second, relaxing the non-negativity condition of LexRSM is highly desirable if we wish to fully unlock the benefit of the lexicographic extension in automated verification. 
A motivating example is given in Fig.~\ref{fig:leadingEx}.
The probabilistic program in Fig.~\ref{fig:leadingEx} terminates almost-surely, but it does not admit any linear GLexRSM (and hence, the GLexRSM synthesis algorithms in~\cite{ChatterjeeGNZZ21} cannot witness its almost-sure termination); 
for example, the function $\boldsymbol{\eta}$ ranks every transition of the program, but violates both leftward and expected leftward non-negativity at the transition $\ell_1 \rightarrow \ell_2$ (note $\boldsymbol{\eta}$ ranks this transition in the third dimension; to check the violation of expected leftward non-negativity, also note $\boldsymbol{\eta}$ ranks $\ell_2 \rightarrow \ell_4$ in the first dimension). 
Here, the source of the problem is that the program has two variables whose progress must be measured
(i.e., increment $y$ to 10 in $\ell_3$; and increment $x$ to 5 in $\ell_4$), but one of their
progress measures can be arbitrarily small during the program execution ($y$ can
be initialized with any value). 
Not only that this structure is rather fundamental, 
it is also 
expected that our desired LexRSM could handle it, 
if it exists. 
Indeed, 
modify the probabilistic program in Fig.~\ref{fig:leadingEx} into a non-probabilistic one 
by changing ``$\mathit{Unif}[1,2]$'' to ``$1$''; then the program admits $\boldsymbol{\eta}$ as a single-component non-negative LexRF.

\medskip
\noindent
{\bf Contributions.}
In this paper, we are the first to
introduce sound LexRSM notions that instantiate single-component non-negative LexRF. 
Our contributions are threefold, as we state below.  

\begin{itemize}
    \item 
First, in response to the first motivation we stated above, we devise a novel notion of \emph{fixability} as a theoretical tool to analyze if 
negative values of a LexRSM ``cause trouble''. 
Roughly speaking, we identify the source of the trouble as ``ill'' exploitation of unbounded negativity of LexRSM;  
our \emph{$\varepsilon$-fixing} operation prohibits such exploitation by basically setting all the negative values of a LexRSM into the same negative value $-\varepsilon$,  
and we say a LexRSM is \emph{$\varepsilon$-fixable} if it retains the ranking condition through such a transformation. 
We give more details about its concept and key ideas in \S\ref{sect:keyideas}. 

The soundness of $\varepsilon$-fixable LexRSM immediately follows from that of strongly non-negative one~\cite{AgrawalCP18} because any LexRSM becomes strongly non-negative through the 
$\varepsilon$-fixing operation (after globally adding $\varepsilon$). Fixable LexRSM instantiates single-component non-negative LexRF for general stochastic processes
(Thm.~\ref{thm:Instantiationfixability}), 
while also serving as a technical basis for proving the soundness of other LexRSMs.   Meanwhile, fixable LexRSM cannot be directly applied to automated verification algorithms due to the inherent non-linearity of $\varepsilon$-fixing;
this observation leads us to our second contribution. 

\item 
Second, 
in response to the second motivation we stated above, 
we introduce \emph{Lazy LexRSM} (LLexRSM) as another LexRSM notion that instantiates single-component non-negative LexRF. 
LLexRSM does not involve the $\varepsilon$-fixing operation in its definition;  
thanks to this property, we have a subclass of LLexRSM that is amenable to automated synthesis via linear programming (see~\S\ref{sect:algorithm}). 
    The LLexRSM condition consists of 
the single-component non-negative LexRSM condition     and     \emph{stability at negativity} we propose     (Def.~\ref{def:LLexRSMmap}),     which roughly requires the following:  
        Once the value of a LexRSM gets negative in some dimension, it must stay negative until that dimension exits the left of the ranking one.     For example, $\boldsymbol{\eta}$ in Fig.~\ref{fig:leadingEx} is an LLexRSM; 
    indeed, $\ell_2 \rightarrow \ell_4$ and $\ell_1 \rightarrow \ell_5$ are the only transitions where     $\boldsymbol{\eta}$ possibly changes its value from negative to non-negative in some dimension (namely, the second one), which is 
    although 
    the right to the ranking dimension (the first one). 
    
    We prove linear LLexRSM is sound for probabilistic programs over linear arithmetics
                (see Thm.~\ref{thm:FairnessImpliesASTofMaps}     for the exact assumption). 
            The proof is highly nontrivial, which is 
    realized by subtle use of a refined variant of fixability; we explain     its core idea in~\S\ref{sect:keyideas}.     Furthermore, Thm.~\ref{thm:FairnessImpliesASTofMaps} shows that expected leftward non-negativity in GLexRSM~\cite{ChatterjeeGNZZ21} is actually redundant under the assumption in Thm.~\ref{thm:FairnessImpliesASTofMaps}. 
    This is surprising, as expected leftward non-negativity has been invented to restore the soundness of leftward non-negative LexRSM, which is generally unsound.
    
    \item 
    Third, we present a synthesis algorithm for the subclass of LLexRSM we mentioned above, and do     experiments; there, our algorithms verified almost-sure termination of     various programs 
    that could not be handled by (a better proxy of) the GLexRSM-based one. The details can be found in~\S\ref{sect:experiment}.
\end{itemize}

\section{Key Observations with Examples}\label{sect:keyideas}
Here we demonstrate by examples how intricate the treatment of negative values of LexRSM is, and how we handle it by our proposed notion of fixability.

\medskip
\noindent
{\bf Blocking ``ill'' exploitation of unbounded negativity.}
Fig.~\ref{fig:CounterEx1} is a counterexample that shows 
leftward non-negative LexRSM is generally unsound 
(conceptually the same as~\cite[Ex. 1]{ChatterjeeGNZZ21}).
The probabilistic program in Fig.~\ref{fig:CounterEx1} does not terminate almost-surely because the chance of entering $\ell_4$ from $\ell_3$ quickly decreases as $t$ increases. 
Meanwhile, $\boldsymbol{\eta}=(\eta_1,\eta_2,\eta_3)$ in Fig.~\ref{fig:CounterEx1} is a leftward non-negative LexRSM over a global invariant $[0 \leq x \leq 1]$; in particular, observe $\eta_2$ decreases by $1$ in expectation from $\ell_3$, whose successor location is either $\ell_4$ or $\ell_1$. 

\input{figures/CounterEx1} 
This example reveals 
an inconsistency between the ways how the single-component non-negativity and ranking condition evaluate the value of a LexRSM, say $\boldsymbol{\eta} = (\eta_1,\ldots,\eta_n)$.
The single-component non-negativity claims $\boldsymbol{\eta}$ cannot rank a transition in a given dimension $k$ whenever $\eta_k$ is negative; 
intuitively, this means that \emph{any} negative value in the ranking domain $\bbR$ should be understood as the same state, namely the ``bottom'' of the domain. 
Meanwhile, the ranking condition evaluates different negative values differently; 
a smaller negative value of $\eta_k$ can contribute more to satisfy the ranking condition, 
as one can see from the behavior of $\eta_2$ in Fig.~\ref{fig:CounterEx1} at $\ell_3$. 
The function $\boldsymbol{\eta}$ in Fig.~\ref{fig:CounterEx1} satisfies the ranking condition over a possibly non-terminating program 
through ``ill'' exploitation of this inconsistency; as $t$ becomes larger,
the value of $\eta_2$ potentially drops more significantly through the transition from $\ell_3$, but with a smaller probability. 

The first variant of our fixability notion, called \emph{$\varepsilon$-fixability}, enables us to ensure that such exploitation is not happening. 
We simply set every negative value in a LexRSM $\boldsymbol{\eta}$ 
to a negative constant $-\varepsilon$, and say $\boldsymbol{\eta}$ is $\varepsilon$-fixable if it retains the ranking condition through the modification\footnote{
To give the key 
ideas in a simpler way, the description here slightly differs from 
the actual definition in \S\ref{sect:fixableInstances}; 
referred results in~\S\ref{sect:keyideas} are derived from the latter.
See Rem.~\ref{rem:diffWithInformalExplanation}. }\label{footnote:difference}. 
For example, the $\varepsilon$-fixing operation changes the value of $\eta_2$ in Fig.~\ref{fig:CounterEx1} at $\ell_4$ from $-2^{t}$ to $-\varepsilon$, 
 and $\boldsymbol{\eta}$ does not satisfy the ranking condition after that. Therefore,  $\boldsymbol{\eta}$ in Fig.~\ref{fig:CounterEx1} is not $\varepsilon$-fixable for any $\varepsilon >0$ (i.e., we successfully reject this $\boldsymbol{\eta}$ through the fixability check). 
   Meanwhile, 
 an $\varepsilon$-fixable LexRSM witnesses almost-sure termination of the underlying program;
 indeed, the fixed LexRSM is a strongly non-negative LexRSM (by  globally adding $\varepsilon$ to the fixed $\boldsymbol{\eta}$), which is known to be sound~\cite{AgrawalCP18}.

The notion of $\varepsilon$-fixability is operationally so simple that one might even feel it is a boring idea; 
nevertheless, 
its contribution to revealing the nature of possibly negative LexRSM is already significant in our paper. Indeed, 
\begin{inparaenum}[(a)]
\item $\varepsilon$-fixable LexRSM instantiates single-component non-negative LexRF with an appropriate $\varepsilon$ (Thm.~\ref{thm:Instantiationfixability});  
\item $\varepsilon$-fixable LexRSM generalizes GLexRSM~\cite{ChatterjeeGNZZ21}, and the proof offers an alternative proof of soundness of GLexRSM that is significantly simpler than the original one (Thm.~\ref{thm:fixabilityGeneralizesGLexRSM}); and
\item its refined variant takes the crucial role in proving soundness of our second LexRSM variant, lazy LexRSM. 
\end{inparaenum}

\medskip
\noindent
{\bf Allowing ``harmless'' unbounded negativity.}
While $\varepsilon$-fixable LexRSM already instantiates single-component non-negative LexRF, 
we go one step further to obtain a LexRSM notion that is amenable to automated synthesis, in particular via \emph{Linear Programming} (LP).
The major obstacle to this end is the case distinction introduced by $\varepsilon$-fixability, which makes the fixed LexRSM nonlinear. 
\emph{Lazy LexRSM} (LLexRSM), our second proposed LexRSM, resolves this problem while it also instantiates 
single-component non-negative LexRF. 

Linear LLexRSM is sound over probabilistic programs with linear arithmetics (Thm.~\ref{thm:FairnessImpliesASTofMaps}). 
The key to the proof is, informally, the following observation: 
\emph{Restrict our attention to probabilistic programs and functions $\boldsymbol{\eta}$ that are allowed in the LP-based synthesis. Then ``ill'' exploitation in Fig.~\ref{fig:CounterEx1} never occurs}, and therefore, a weaker condition than $\varepsilon$-fixability (namely, the LLexRSM one) suffices for witnessing program termination. 
In fact, Fig.~\ref{fig:CounterEx1}
involves 
\begin{inparaenum}[(a)]
\item non-linear arithmetics in the program, 
\item parametrized if-branch in the program (i.e., the grammar `` {\bf if prob}$(p)$ {\bf then} $P$ {\bf else} $Q$ {\bf fi} '' with $p$ being a variable), and
\item non-linearity of $\boldsymbol{\eta}$. 
\end{inparaenum}
None of them are allowed in the LP-based synthesis (at least, in the standard LP-based synthesis via Farkas' Lemma~\cite{chakarov2013probabilistic, AgrawalCP18, ChatterjeeGNZZ21}). 
Our informal statement above is formalized as Thm.~\ref{thm:fixabilityOfLLexRSM}, which roughly says: 
Under such a restriction to probabilistic programs and $\boldsymbol{\eta}$, any LLexRSM 
is \emph{$(\varepsilon, \gamma)$-fixable}.  
Here, $(\varepsilon, \gamma)$-fixability is a refined version of $\varepsilon$-fixability; 
while it also ensures that ``ill'' exploitation is not happening in $\boldsymbol{\eta}$, it is less restrictive than $\varepsilon$-fixability by allowing ``harmless'' unbounded negative values of $\boldsymbol{\eta}$. 

\input{figures/CounterEx2}
Fig.~\ref{fig:CounterEx2} gives an example of such a harmless behavior of $\boldsymbol{\eta}$ rejected by $\varepsilon$-fixability. 
It also shows why we cannot simply use $\varepsilon$-fixability to check  
an LLexRSM does not do ``ill'' exploitation.
The function $\boldsymbol{\eta}= (\eta_1, \eta_2)$ in Fig.~\ref{fig:CounterEx2} is leftward non-negative over the global invariant $[0 \leq x \leq 1 \land t\geq 1]$, so it is an LLexRSM for the probabilistic program there;  
the program and $\boldsymbol{\eta}$ are also in the scope of LP-based synthesis; but $\boldsymbol{\eta}$ is not $\varepsilon$-fixable for any $\varepsilon>0$.
Indeed, the $\varepsilon$-fixing operation changes the value of $\eta_2$ at $\ell_4$ from $-2t-4$ to $-\varepsilon$, and $\boldsymbol{\eta}$ does not satisfy the ranking condition at $\ell_2$ after the change. 
Here we notice that, however, 
the unbounded negative values of $\eta_2$ are ``harmless'';  
that is, the ``ill-gotten gains'' by the unbounded negative values of $\eta_2$ at $\ell_4$ are only ``wasted'' to unnecessarily increase $\eta_2$ at $\ell_3$. 
In fact, 
$\boldsymbol{\eta}$ still satisfies the ranking condition if we change the value of $\eta_2$ at $\ell_1,\ell_2,\ell_3$ to $2,1$, and $0$, respectively. 

We resolve this issue by partially waiving the ranking condition of $\boldsymbol{\eta}$ after the $\varepsilon$-fixing operation. 
It is intuitively clear that the program in Fig.~\ref{fig:CounterEx2} almost-surely terminates, 
and the intuition here is that the program essentially repeats an unbiased coin tossing 
until the tail is observed (here, ``observe the tail'' corresponds to ``observe $\mbox{\bf prob}(0.5) = \mbox{\bf true}$ at $\ell_2$''). 
This example tells us that, 
to witness the almost-sure termination of this program, we only need to guarantee the program (almost-surely) visits either the terminal location $\ell_5$ \emph{or the ``coin-tossing location'' $\ell_2$} from anywhere else. The $\varepsilon$-fixed $\boldsymbol{\eta}$ in Fig.~\ref{fig:CounterEx2} does witness such a property of the program, as it ranks every transition \emph{except those that are from a coin-tossing location}, namely $\ell_2$.

We generalize this idea as follows: 
Fix $\gamma \in (0,1)$, and say a program state is a ``coin-tossing state'' for $\boldsymbol{\eta} = (\eta_1, \ldots, \eta_n)$ in the $k$-th dimension if $\eta_k$ drops from non-negative to negative (i.e., the ranking is ``done'' in the $k$-th dimension) with the probability $\gamma$ or higher.
Then we say $\boldsymbol{\eta}$ is \emph{$(\varepsilon, \gamma)$-fixable} (Def.~\ref{def:egfixability}) if the $\varepsilon$-fixed $\boldsymbol{\eta}$ 
is a strongly non-negative LexRSM (after adding $\varepsilon$) \emph{except that}, 
at each coin-tossing state, we waive the ranking condition of $\boldsymbol{\eta}$ in the corresponding dimension.
For example, $\boldsymbol{\eta}$ in Fig.~\ref{fig:CounterEx2} is $(\varepsilon, \gamma)$-fixable for any $\gamma \in (0, 0.5]$. 
As expected,  $(\varepsilon, \gamma)$-fixable 
 LexRSM  is sound  for any $\varepsilon>0$ and $\gamma \in (0,1)$ 
 (Cor.~\ref{cor:ecfixableWitnessAST}).

\section{Preliminaries}\label{sect:preliminaries}
We recall the technical 
preliminaries. 
Omitted details are 
\begin{arxiv}
    in Appendix~\ref{append:preliminaries}. 
\end{arxiv}
\begin{cav}
    in~\cite[Appendix A]{TakisakaZWL24arXiv}.
\end{cav}

\medskip
\noindent
{\bf Notations.} 
We assume the readers are familiar with the basic notions of measure theory, see e.g.~\cite{Ash:book,BertsekasS07}. 
The sets of non-negative integers and reals are denoted by $\bbN$ and $\bbR$, respectively. 
The collection of all Borel sets of a topological space $\mathcal{X}$ is denoted by $\calB(\mathcal{X})$. 
The set of all probability distributions over the measurable space $(\Omega,\calB(\Omega))$ is denoted by $\calD(\Omega)$. 
The value of a vector $\boldsymbol{x}$ at the $i$-th index is denoted by $\boldsymbol{x}[i]$ or $x_i$. 
A subset $D \subseteq \bbR$ of the reals is \emph{bounded} if $D \subseteq [-x,x]$ for some $x>0$.

For a finite variable set $V$ and the set $val^V$ of its valuations, we form \emph{predicates} as first-order formulas with atomic predicates of the form $f \leq g$, 
where $f,g\colon val^V \to R$ and $R$ is linearly ordered. 
Often, we are only interested in the value of a predicate $\varphi$ over a certain subset $\mathcal{X} \subseteq val^V$, in which case, we call $\varphi$ a predicate \emph{over $\mathcal{X}$}. 
We identify a predicate $\varphi$ over $\mathcal{X}$ with a function $\tilde\varphi\colon \mathcal{X} \to \br{0,1}$ such that $\tilde\varphi(x)=1$ if and only if $\varphi(x)$ is true. 
The \emph{semantics} of $\varphi$, i.e., the set $\{x\in \mathcal{X} \mid \varphi(x) \mbox{ is true}\}$, is denoted by $\llbracket \varphi \rrbracket$. The \emph{characteristic function} $\mychar_A: \mathcal{X} \to \{0,1\}$ of a subset $A$ of $\mathcal{X}$ is a function such that $\llbracket \mychar_A =1 \rrbracket = A$. 
For a probability space $(\Omega,\calF,\bbP)$, 
we say $\varphi$ over $\Omega$ is \emph{($\calF$-)measurable} when $\sem{\varphi}\in \calF$.
For such a $\varphi$, the \emph{satisfaction probability} of $\varphi$ w.r.t. $\bbP$, i.e., the value $\bbP(\llbracket \varphi \rrbracket)$, is also denoted by $\bbP(\varphi)$; 
we say \emph{$\varphi$ holds $\bbP$-almost surely} ($\bbP$-a.s.) if $\bbP(\varphi) = 1$.

\subsection{Syntax and Semantics of Probabilistic Programs}\label{sect:SyntaxSemantics}
{\bf Syntax.} 
We define the syntax of \emph{Probabilistic Programs} (PPs) 
similarly to e.g.,~\cite{AgrawalCP18,TakisakaOUH21}. More concretely, 
PPs have the standard control structure in imperative languages such as 
if-branches and while-loops, while the if-branching and variable assignments can also be done in either  nondeterministic or probabilistic ways. Namely, 
$\mbox{`\textbf{if}} \, \mbox{$\star$'}$ describes a nondeterministic branching;  
$ \mbox{`\textbf{ndet$(D)$}'}$ describes a nondeterministic assignment chosen from a bounded\footnote{
This is also assumed in~\cite{ChatterjeeGNZZ21}
to avoid a complication 
in possibly negative LexRSMs. 
} domain $D \subseteq \calB(\bbR)$; 
$\mbox{`\textbf{if}} \, \mbox{\textbf{prob$(p)$}'}$ 
with a constant $p \in [0,1]$
describes a probabilistic branching that executes the 
$\mbox{`\textbf{then}'}$ branch with probability $p$, or the 
$\mbox{`\textbf{else}'}$ branch with probability $1-p$; and 
$\mbox{`\textbf{sample$(d)$}'}$ describes a probabilistic assignment sampled from a distribution $d\in \calD(\bbR)$. 
We consider PPs without \emph{conditioning}, which are also called \emph{randomized programs}~\cite{TakisakaOUH21}; PPs with conditioning are considered in e.g.~\cite{OlmedoGJKKM18}. 
The exact grammar is given 
\begin{arxiv}
    in Appendix~\ref{append:preliminaries}. 
\end{arxiv}
\begin{cav}
    in~\cite[Appendix A]{TakisakaZWL24arXiv}.
\end{cav}

In this paper, we focus our attention on PPs with linear arithmetics; 
we say a PP is \emph{linear} if each arithmetic expression in it is 
linear, i.e., of the form $b+\sum_{i=1}^n a_i \cdot v_i$ for constants $a_1,\ldots, a_n, b$ and program variables $v_1,\ldots, v_n$.

\medskip
\noindent
{\bf Semantics.}
We adopt \emph{probabilistic control flow graph} (pCFG) as the semantics of PPs, which is standard in existing RSM works (e.g., \cite{chakarov2013probabilistic,TakisakaOUH21,ChatterjeeGNZZ21}). Informally, it is a labeled directed graph whose vertices are program locations, 
and whose edges represent possible one-step executions in the program. 
Edges are labeled with the necessary information so that one can reconstruct the PP represented by the pCFG; 
for example, an edge $e$ can be labeled with 
the assignment commands executed through $e$ (e.g., `$x:=x+1$'), 
the probability $p\in[0,1]$ that $e$ will be chosen (through  $\mbox{`\textbf{if}} \, \mbox{\textbf{prob$(p)$}'}$), 
the guard condition, and so on. 
Below we give its formal definition for completeness; 
see 
\begin{arxiv}
    Appendix~\ref{append:preliminaries} 
\end{arxiv}
\begin{cav}
    \cite[Appendix A]{TakisakaZWL24arXiv}
\end{cav}
for how to translate PPs into pCFGs.

\begin{mydefinition}[pCFG]
A \emph{pCFG} is a tuple
$(L, V, \Delta, \myup, G)$, 
where
\begin{enumerate}
 \item $L$ is a finite set of locations.
 \item $V = \{x_1, \ldots , x_{|V|}\}$ is a finite set of program variables.
 \item $\Delta$ is a finite set of \emph{(generalized) transitions}\footnotemark, i.e., tuples $\tau  = (\ell,\delta)$ of a location $\ell\in L$ and a distribution $\delta\in \calD(L)$ over successor locations.
 \item $\myup$ is a function that receives a transition $\tau\in \Delta$ and returns a tuple $(i,u)$  of 
 a target variable index $i \in \br{1,\ldots,|V|}$ and an update element $u$.  
 Here, $u$ is either
\begin{inparaenum}
\item a Borel measurable function $u: \bbR^{|V|} \to \bbR$,
\item a distribution $d \in \calD(\bbR)$, or
\item a bounded measurable set $R \in \calB(\bbR)$. 
In each case, we say $\tau$ is \emph{deterministic}, \emph{probabilistic}, and \emph{non-deterministic}, respectively; 
the collections of these transitions are denoted by $\Delta_d$, $\Delta_p$, and $\Delta_n$, respectively. 
\end{inparaenum}
 \item $G$ is a guard function that assigns a  $G(\tau):\bbR^{|V|} \to \br{0,1}$ to each  $\tau \in \Delta$.
\end{enumerate}
\footnotetext{Defining these as edges might be more typical, as in our informal explanation. We adopt the style of~\cite{AgrawalCP18,ChatterjeeGNZZ21} for convenience; it can handle $\mbox{`\textbf{if}} \, \mbox{\textbf{prob$(p)$}'}$ by a single $\tau$. }
\end{mydefinition}

Below we fix a pCFG $\calC = (L, V, \Delta, \myup, G)$. 
A \emph{state} of $\calC$ is a tuple $s = (\ell,\boldsymbol{x})$ of location $\ell \in L$ and variable assignment vector $\boldsymbol{x} \in \bbR^{|V|}$. 
We write $\calS$ to denote the state set $L \times \bbR^{|V|}$. 
Slightly abusing the notation, for $\tau = (\ell, \delta)$, we identify the set $\sem{G(\tau)} \subseteq \bbR^{|V|}$ and the set $\{\ell\}\times \sem{G(\tau)} \subseteq \calS$; in particular, we write $s\in \sem{G(\tau)}$ when $\tau$ is \emph{enabled at $s$}, i.e., $s=(\ell,\boldsymbol{x})$, $\tau =(\ell, \delta)$ and $\boldsymbol{x}\in \sem{G(\tau)}$.

A pCFG $\calC$ with its state set $\calS$ can be understood as a transition system over $\calS$ with probabilistic transitions and nondeterminism (or, more specifically, a Markov decision process with its states $\calS$). 
Standard notions such as \emph{successors} of a state $s \in \calS$, \emph{finite paths}, and \emph{(infinite) runs} of $\calC$ 
are defined 
as the ones over 
such a transition system. 
The set of all successors of $s \in \sem{G(\tau)}$ via $\tau$ is denoted by $\mysucc_\tau(s)$. 
The set of runs of $\calC$ is denoted by $\Pi_\calC$. 

\emph{Schedulers} resolve nondeterminism in pCFGs. Observe there are two types of nondeterminism: \begin{inparaenum}[(a)]
    \item nondeterministic choice of $\tau \in \Delta$ at a given state (corresponds to $\mbox{`\textbf{if}} \, \mbox{$\star$'}$), and \label{item:ndet1}
    \item nondeterministic variable update in a nondeterministic transition $\tau \in \Delta_n$ (corresponds to $ \mbox{`$x_i:=$\textbf{ndet$(D)$}'}$). \label{item:ndet2}
\end{inparaenum}
We say a scheduler is \emph{$\Delta$-deterministic} if its choice is non-probabilistic in Case (\ref{item:ndet1}).

We assume pCFGs are deadlock-free; 
we also assume that there are designated locations $\ell_{\mathrm{in}}$ and $\ell_{\mathrm{out}}$ that represent program initiation and termination, respectively.
An \emph{initial state} is a state of the form $(\ell_{\mathrm{in}}, \boldsymbol{x})$. 
We assume a transition from $\ell_{\mathrm{out}}$ is unique, denoted by $\tau_{\mathrm{out}}$; this transition does not update anything.

By fixing a scheduler $\sigma$ and an initial state $s_I$, 
the infinite-horizon behavior of $\calC$ is determined  
as a distribution $\bbP_{s_I}^\sigma$ over $\Pi_\calC$; that is, for a measurable $A\subseteq \Pi_\calC$, the value 
$\bbP_{s_I}^\sigma(A)$  
is the probability that a run of $\calC$ from $s_I$ is in $A$ under $\sigma$. We call the probability space $(\Pi_\calC, \calB(\Pi_\calC), \bbP_{s_I}^\sigma)$ the \emph{dynamics of $\calC$ under $\sigma$ and $s_I$}. 
See~\cite{BertsekasS07} for the formal construction; a brief explanation is 
\begin{arxiv}
    in Appendix~\ref{append:preliminaries}.
\end{arxiv}
\begin{cav}
    in~\cite[Appendix A]{TakisakaZWL24arXiv}.
\end{cav}

We define the \emph{termination time} of a pCFG $\calC$ as the function $T_{\mathrm{term}}^\calC:\Pi_\calC \to \bbN \cup \{+\infty\}$ such that 
$T_{\mathrm{term}}^\calC(s_0s_1\ldots) = \inf\br{t \in \bbN \mid \exists \boldsymbol{x}. s_t = (\ell_{\mathrm{out}}, \boldsymbol{x})}$. 
Now we formalize our objective, i.e., 
almost-sure termination of pCFG, 
as follows.

\begin{mydefinition}[AST of pCFG]\label{def:ASTermPCFG}
    A run $\omega \in \Pi_\calC$ \emph{terminates} if $T_{\mathrm{term}}^\calC(\omega) <\infty$. 
    A pCFG $\calC$ is \emph{a.s. terminating} (AST) under a scheduler     $\sigma$ and an initial state $s_I$ if a run of $\calC$ terminates $\bbP_{s_I}^\sigma$-a.s.     We say $\calC$ is AST if it is AST for any $\sigma$ and $s_I$.
\end{mydefinition}

\subsection{Lexicographic Ranking Supermartingales}\label{subsect:LexRSM}
Here we recall mathematical preliminaries of the LexRSM theory. 
A (Lex)RSM typically comes in two different forms: 
one is a vector-valued function $\boldsymbol{\eta}: \calS \to \bbR^n$ over states $\calS$ of a pCFG $\calC$, and 
another is a stochastic process over the runs $\Pi_\calC$ of $\calC$. We recall relevant notions in these formulations, which are frequently used in existing RSM works~\cite{chakarov2013probabilistic,ChatterjeeGNZZ21}. 
We also recall the formal definition of LexRSMs with three different non-negativity conditions in Fig.~\ref{fig:ExForGLexRSM}.

\medskip\noindent{\bf LexRSM as a quantitative predicate.}
Fix a pCFG $\calC$. 
An \emph{($n$-dimensional) measurable map} (MM) is a Borel measurable function $\boldsymbol{\eta}:\calS \to \bbR^n$.
For a given 1-dimensional MM $\eta$ and a transition $\tau$, 
The (maximal) \emph{pre-expectation} of $\eta$ under  $\tau$ is a function that formalizes ``the value of $\eta$ after the transition $\tau$''. 
More concretely, 
it is a function 
$\overline{\bbX}_\tau \eta:\sem{G(\tau)} \to \bbR$ that returns, for a given state $s$, the maximal expected value of $\eta$ at the successor state of $s$ via $\tau$. 
Here, the maximality refers to the set of all possible nondeterministic choices at $s$.

A \emph{level map} $\Lv:\Delta \to \br{0,\ldots, n}$ designates the ranking dimension of 
the associated 
LexRSM $\boldsymbol{\eta}: \calS \to \bbR^n$. 
We require $\Lv(\tau) = 0$ if and only if $\tau = \tau_{\mathrm{out}}$. 
We say an MM $\boldsymbol{\eta}$ \emph{ranks} a transition $\tau$ in the dimension $k$ (under $\Lv$) when $k = \Lv(\tau)$. 
An \emph{invariant} is a measurable predicate $I: \calS \to \br{0,1}$ such that 
$\sem{I}$ is closed under transitions and 
$\ell_{\mathrm{in}} \times \bbR^{|V|} \subseteq \sem{I}$. The set $\sem{I}$ over-approximates the reachable states in $\calC$.

Suppose an $n$-dimensional MM $\boldsymbol{\eta}$ and an associated level map $\Lv$ are given. 
We say $\boldsymbol{\eta}$ satisfies \emph{the ranking condition} (under $\Lv$ and $I$) if 
the following holds for each 
$\tau \neq \tau_{\mathrm{out}}$, 
$ s \in \sem{I \land G(\tau)}$, and 
$k \in \br{1,\ldots, \Lv(\tau)}$: \begin{align*}
    \overline\bbX_\tau\boldsymbol{\eta}[k](s)\leq
    \begin{cases}
        \boldsymbol{\eta}[k](s)        & \text{if } k <  \Lv(\tau),\\  
        \boldsymbol{\eta}[k](s) -1    & \text{if } k =  \Lv(\tau).
    \end{cases}
\end{align*}
We also define the three different non-negativity conditions in Fig.~\ref{fig:ExForGLexRSM}, i.e., \emph{STrong (ST)}, \emph{LeftWard (LW)}, and \emph{Single-Component (SC)} non-negativity, as follows: 
\begin{align*}
    \mbox{(ST non-neg.) } 
      & \forall s \in \sem{I}. 
        \forall k \in \br{1,\ldots, n} .&&
        \boldsymbol{\eta}[k](s) \geq 0,\\      
    \mbox{(LW non-neg.) } 
      & \forall\tau \neq \tau_{\mathrm{out}}. 
        \forall s \in \sem{I \land G(\tau)}. 
        \forall k \in \br{1,\ldots, \Lv(\tau)}.&&
        \boldsymbol{\eta}[k](s) \geq 0,\\    
    \mbox{(SC non-neg.) } 
      & \forall\tau \neq \tau_{\mathrm{out}}. 
        \forall s \in \sem{I \land G(\tau)}. &&
        \boldsymbol{\eta}[\Lv(\tau) ](s) \geq 0.
\end{align*}

All the materials above are wrapped up in the following definition.

\begin{mydefinition}[(ST/LW/SC)-LexRSM map]\label{def:SC-LexRSMMap}
Fix a pCFG $\calC$ with an invariant $I$. 
Let $\boldsymbol{\eta}$ be an MM associated with a level map $\Lv$. 
The MM $\boldsymbol{\eta}$ is called a \emph{STrongly non-negative 
LexRSM 
map (ST-LexRSM map)} over $\calC$ supported by $I$ if 
it 
satisfies the ranking condition and the strong non-negativity under $\Lv$ and $I$. 
If it satisfies the leftward or single-component non-negativity instead of the strong one, then we call it \emph{LW-LexRSM map} or \emph{SC-LexRSM map}, respectively.
\end{mydefinition}
\noindent{\bf LexRSM as a stochastic process.}
When it comes to automated synthesis, a (Lex)RSM is usually a function $\boldsymbol{\eta}$ over program states, as defined in Def.~\ref{def:SC-LexRSMMap}. 
Meanwhile, when we prove the properties of (Lex)RSMs themselves (e.g., soundness), it is often necessary to inspect the behavior of $\boldsymbol{\eta}$ upon the program execution under given scheduler $\sigma$ and initial state $s_I$.  
Such a behavior of $\boldsymbol{\eta}$ is formalized as a sequence $(\bfX_t)_{t=0}^\infty$ of random variables  over the dynamics of the underlying pCFG, which forms a \emph{stochastic process}. 

A (discrete-time) \emph{stochastic process} in a probability space $(\Omega, \calF, \bbP)$ is a sequence $(\bfX_t)_{t=0}^\infty$ of $\calF$-measurable random variables $\bfX_t: \Omega \to \bbR^n$ for $t \in \bbN$. 
In our context, it
is typically associated with another random variable $T:\Omega \to \bbN \cup \{+\infty\}$ that describes the termination time of $\omega\in \Omega$.  
We say $T$ is \emph{AST (w.r.t. $\bbP$)} if $\bbP(T <\infty) = 1$; 
observe that, if $(\Omega, \calF, \bbP)$ is the dynamics of a pCFG $\calC$ under $\sigma$ and $s_I$, then $\calC$ is AST under $\sigma$ and $s_I$ if and only if $T_{\mathrm{term}}^\calC$ is AST w.r.t. $\bbP$.
As standard technical requirements, we assume there is a \emph{filtration} $(\calF_t)_{t=0}^\infty$ in $(\Omega, \calF, \bbP)$ such that 
$(\bfX_t)_{t=0}^\infty$ is \emph{adapted} to $(\calF_t)_{t=0}^\infty$, $T$ is a \emph{stopping time} w.r.t.
$(\calF_t)_{t=0}^\infty$, and $(\bfX_t)_{t=0}^\infty$ is \emph{stopped} at $T$; 
\begin{arxiv}
    see Appendix~\ref{append:preliminaries}
\end{arxiv}
\begin{cav}
    see~\cite[Appendix A]{TakisakaZWL24arXiv}
\end{cav}
for their definitions.

For a stopping time $T$ w.r.t. $(\calF_t)_{t=0}^\infty$, we define a \emph{level map} $\seqOf{\Lv_t}$ as a sequence of $\calF_t$-measurable functions 
$\Lv_t: \Omega \to \br{0,\ldots n}$ such that $\sem{\Lv_t=0} = \sem{T\leq t}$ for each $t$. 
We call a pair of a stochastic process and a level map an \emph{instance for $T$}; 
just like we construct an MM $\boldsymbol{\eta}$ and a level map $\Lv$ as an AST certificate of a pCFG $\calC$, we construct an instance for a stopping time $T$ as its AST certificate. 
We say an instance $(\seqOf{\bfX_t},\seqOf{\Lv_t})$ for $T$ \emph{ranks} $\omega\in\Omega$ in the dimension $k$ at time $t$ when $T(\omega) > t$ and  $k = \Lv_t(\omega)$. 

For $c>0$, we say an instance $(\seqOf{\bfX_t},\seqOf{\Lv_t})$ satisfies the \emph{$c$-ranking condition} if, for each 
$t \in \bbN$, $\omega \in \sem{\Lv_t \neq 0}$, and 
$k \in \br{1, \ldots, \Lv_t(\omega)}$,
we have:
\begin{align}
        \bbE[\bfX_{t+1}[k] \mid \calF_t] (\omega) \leq 
        \bfX_t[k](\omega) -c \cdot \mychar_{\sem{k=\Lv_t }}(\omega) \quad (\bbP\mbox{-a.s.})\label{eq:rankingIneq}
\end{align}
Here, the function $\bbE[\bfX_{t+1}[k] \mid \calF_t]$ denotes the \emph{conditional expectation} of $\bfX_{t+1}[k]$ given $\calF_t$, which takes the role of pre-expectation. We mostly let $c=1$ and simply call it the ranking condition; 
the only result sensitive to $c$ is Thm.~\ref{thm:Instantiationfixability}.

We also define the three different non-negativity conditions for an instance as follows. 
Here we adopt a slightly general (but essentially the same) variant of strong non-negativity instead, calling it \emph{uniform well-foundedness}; 
we simply allow the uniform lower bound to be any constant $\bot \in \bbR$ instead of fixing it to be zero. 
This makes the later argument simpler. 
\begin{align*}
    \mbox{(UN well-fnd.) } 
      & \exists \bot \in \bbR.
        \forall t \in \bbN. 
        \forall \omega \in \Omega.         \forall k \in \br{1, \ldots, n}.&&
        \hspace{-0.1cm}
        \bfX_t[k](\omega) \geq \bot, \\
    \mbox{(LW non-neg.) } 
      & \forall t \in \bbN. 
        \forall \omega \in 
        \sem{\Lv_t \neq 0}.
                \forall k \in \br{1, \ldots, \Lv_t(\omega)}.&&
        \hspace{-0.1cm}
        \bfX_t[k](\omega) \geq 0, \\
    \mbox{(SC non-neg.) } 
      & \forall t \in \bbN. 
        \forall \omega \in \sem{\Lv_t \neq 0}. &&
                                \hspace{-0.1cm}
        \bfX_t[\Lv_t(\omega)](\omega) \geq 0.
\end{align*}

\begin{mydefinition}[(UN/LW/SC)-LexRSM]\label{def:LexRSMs}
Suppose the following are given:
a probability space $(\Omega, \calF, \bbP)$;
a filtration $(\calF_t)_{t=0}^\infty$ on $\calF$; and 
a stopping time $T$ w.r.t. $(\calF_t)_{t=0}^\infty$.
An instance $\calI = (\seqOf{\bfX_t},\seqOf{\Lv_t})$ 
is called 
a \emph{UNiformly well-founded LexRSM (UN-LexRSM)} for $T$ with the bottom $\bot \in \bbR$ 
and a constant $c \in \bbR$
if \begin{inparaenum}[(a)]
    \item $\seqOf{\bfX_t}$ is adapted to $\seqOf{\calF_t}$;\label{item:SC-LexRSMAdaptedness}
    \item for each $t \in \bbN$ and $1 \leq k \leq n$, the expectation 
    of $\bfX_t[k]$
        exists; \label{item:SC-LexRSMExistenceOfExpectation}
    \item $\calI$ satisfies the $c$-ranking condition; and\label{item:SC-LexRSMRanking}
    \item $\calI$ is uniformly well-founded with the bottom $\bot$.\label{item:SC-LexRSMNonneg}
\end{inparaenum}
We define \emph{LW-LexRSM} and \emph{SC-LexRSM} by changing (\ref{item:SC-LexRSMNonneg}) with 
LW and SC non-negativity, respectively.
\end{mydefinition}

We mostly assume $c=1$ and omit to mention the constant.
UN-LexRSM is known to be sound~\cite{AgrawalCP18}; meanwhile, LW and SC-LexRSM are generally unsound~\cite{ChatterjeeGNZZ21,ferrer2015probabilistic}. 
We still mention the latter two as parts of sound LexRSMs.

\medskip\noindent{\bf From RSM maps to RSMs.} 
Let $\boldsymbol{\eta}$  be an MM over a pCFG $\calC$ with a level map 
$\Lv$. 
Together with a $\Delta$-deterministic
scheduler $\sigma$ and initial state $s_I$, 
it \emph{induces} 
an instance $(\seqOf{\bfX_t},\seqOf{\Lv_t})$ 
over the dynamics of $\calC$,  by letting $\bfX_t(s_0s_1\ldots) = \boldsymbol{\eta}(s_t)$;
it describes the behavior of $\boldsymbol{\eta}$ and $\Lv$ 
through executing $\calC$ from $s_I$ under $\sigma$. 
Properties of $\boldsymbol{\eta}$ such as ranking condition or non-negativity are inherited to the induced instance (if the expectation of $\bfX_t[k]$ 
exists for each $t$, $k$). 
For example,
an instance induced by an ST-LexRSM map is an UN-LexRSM with $\bot = 0$.

\medskip\noindent{\bf Non-probabilistic settings, and instantiation of SC-LexRF.} 
The key question \todo{can be shortend if necessary} in this paper is to find a 
LexRSM notion that instantiates SC non-negative LexRF (or SC-LexRF for short); 
that is, we would like to find a LexRSM notion whose conditions are satisfied by 
\begin{arxiv}
SC-LexRSM\footnote{
One would perhaps expect to see ``SC-Lex\emph{RF}'' here; such a change does not make a difference under a canonical definition of SC-LexRF, so we define the notion of instantiation in this way to save space. See also 
    Appendix~\ref{append:preliminaries}. 
    }\label{footnote:instantiation}
\end{arxiv}
\begin{cav}
SC-LexRSM\footnote{
One would perhaps expect to see ``SC-Lex\emph{RF}'' here; such a change does not make a difference under a canonical definition of SC-LexRF, so we define the notion of instantiation in this way to save space. See also~\cite[Appendix A]{TakisakaZWL24arXiv}.
    }
\end{cav}
in the \emph{non-probabilistic setting}, which we formalize as follows. 
We say a pCFG is a \emph{(non-probabilistic) CFG} if 
\begin{inparaenum}[(a)]
    \item $\delta$ is Dirac
    for each $(\ell, \delta) \in \Delta$,     and
    \item $\Delta_p = \emptyset$; 
\end{inparaenum} 
this roughly means that a CFG is 
  a model of a PP without $\mbox{`\textbf{if}} \, \mbox{\textbf{prob$(p)$}'}$ and $\mbox{`\textbf{sample$(d)$}'}$. 
We say a probability space $(\Omega, \calF, \bbP)$ is \emph{trivial} if $\Omega$ is a singleton, say $\br{\omega}$.

\begin{full}
    Let $\boldsymbol{\eta}:S \to \bbR^n$ be an MM with $\Lv: \Delta \to \br{0,\ldots,n}$. 
For $\Delta$-deterministic\footnote{
    This is relevant for defining $\Lv_t$.     Similar RSM construction for general $\sigma$ is also possible by defining runs of pCFGs as alternating sequences of states and transitions.
}
scheduler $\sigma$ and initial state $s_I$, 
we have a canonical stochastic process induced by them. 
Concretely, we say an instance $(\seqOf{\bfX_t}, \seqOf{\Lv_t})$ is \emph{induced} by $\boldsymbol{\eta}$ and $\Lv$ under $\sigma$ and $s_I$ if it is constructed as follows:
the underlying probability space $(\Omega, \calF, \bbP)$ is the dynamics of $\calC$ under $\sigma$ and $s_I$;
the underlying filtration $\seqOf{\calF_t}$ is the one generated by cylinders of length $t$;
the stochastic process $\seqOf{\bfX_t}$ is as in Eq. (\ref{eq:SPofMM}); 
the stopping time $T$ is the termination time of $\calC$; 
the level map is defined by $\Lv_t(c_0c_1\ldots) = \Lv(\sigma_\Delta(c_0\ldots c_t))$.

If $\boldsymbol{\eta}$ is an SC-LexRSM map with $\Lv$, then $(\seqOf{\bfX_t}), \seqOf{\Lv_t})$ is an SC-LexRSM provided that relevant values in the SC-LexRSM constraint is well-defined. 
Concretely, we say an MM $\boldsymbol{\eta}$ is \emph{adequate (over $\calC$)} if
$\bbE[\bfX_t[k]]$ is well-defined for each $t\in \bbN$, $k \in \br{1,\ldots,n}$, and any $\sigma$ and $s_I$ that induces $(\seqOf{\bfX_t})$.

\begin{myproposition}[\cite{ChatterjeeGNZZ21}]
Let $\boldsymbol\eta$ be an adequate $n$-dimensional SC-LexRSM map with a level map $\Lv:\Delta \to \br{0,\ldots,n}$. 
Let $(\seqOf{\bfX_t}), \seqOf{\Lv_t})$ be induced by $\boldsymbol{\eta}$ and $\Lv$ under $\sigma$ and $r_0$.
Then $((\bfX_t)_{t=0}^\infty, (\Lv_t)_{t=0}^\infty)$ is an SC-LexRSM. \end{myproposition}
\end{full}

\section{Fixable LexRSMs}\label{sect:fixableInstances}
In \S\ref{sect:fixableInstances}-\ref{sect:algorithm} we give our novel technical notions and results. 
In this section, we will introduce the notion of fixability and related results. 
Here we focus on technical rigorousness and conciseness, see~\S\ref{sect:keyideas} for the underlying intuition. Proofs are given in 
\begin{arxiv}
    appendices. 
\end{arxiv}
\begin{cav}
    appendices of~\cite{TakisakaZWL24arXiv}.
\end{cav}
We begin with the formal definition of $\varepsilon$-fixability. 

\begin{myremark}\label{rem:diffWithInformalExplanation}
As in Footnote~\ref{footnote:difference}, our formal definitions of fixability in this section slightly differ 
from an informal explanation in~\S\ref{sect:keyideas}. 
One difference is that the $\varepsilon$-fixing in Def.~\ref{def:epsilonfixing} changes the value of a LexRSM at dimension $k$ whenever it is negative \emph{or} $k$ is strictly on the right to the ranking dimension. 
This modification is necessary to prove Thm.~\ref{thm:fixabilityGeneralizesGLexRSM}. 
Another is that we define fixability as the notion for an instance $\calI$, rather than for an MM $\boldsymbol{\eta}$. 
While the latter can be also done in an obvious way (as informally done in \S\ref{sect:keyideas}), we do not formally do that because it is not necessary for our technical development. 
One can ``fix'' the argument in \S\ref{sect:keyideas} into 
the one over instances 
by translating ``fixability of $\boldsymbol{\eta}$'' to ``fixability of an instance induced by $\boldsymbol{\eta}$''.

\end{myremark}

\begin{mydefinition}[$\varepsilon$-fixing of an instance]\label{def:epsilonfixing}
  Let $\calI =(\seqOf{\bfX_t}), \seqOf{\Lv_t})$ be an instance for a stopping time $T$, and let $\varepsilon > 0$.  
  The \emph{$\varepsilon$-fixing} of $\calI$ is another instance $\tilde{\calI} = (\seqOf{\tilde{\bfX}_t},\seqOf{\Lv_t})$ for 
    $T$, where
\begin{align*}
\tilde\bfX_t[k](\omega) = 
    \begin{cases}
        -\varepsilon         & \text{if } \bfX_t[k](\omega)  < 0 \text{ or } k > \Lv_t(\omega),\\         \bfX_t[k](\omega)    & \text{otherwise}.    \end{cases}
\end{align*}
  We say an SC-LexRSM 
  $\calI$ is \emph{$\varepsilon$-fixable}, or call it an \emph{$\varepsilon$-fixable LexRSM}, if its $\varepsilon$-fixing $\tilde{\calI}$
  is an UN-LexRSM with the bottom $\bot = -\varepsilon$.
  \end{mydefinition}

Observe that the $\varepsilon$-fixing of any instance is uniformly well-founded with the bottom $\bot = -\varepsilon$, so the $\varepsilon$-fixability only asks if the ranking condition is preserved through $\varepsilon$-fixing. 
Also, observe that the soundness of $\varepsilon$-fixable LexRSM immediately follows from that of UN-LexRSM~\cite{AgrawalCP18}. 

While we do not directly use $\varepsilon$-fixability as a technical tool, 
the two theorems below show its conceptual value. 
The first one answers 
our key problem: $\varepsilon$-fixable LexRSM
instantiates SC-LexRF with sufficiently large $\varepsilon$.

\begin{mytheorem}[fixable LexRSM instantiates SC-LexRF]\label{thm:Instantiationfixability}
        Suppose $\calI = (\seqOf{\boldsymbol{x}_t},  \seqOf{\Lv_t})$ is  
    an SC-LexRSM for a stopping time $T$ over the trivial probability space with a constant $c$, and let $\varepsilon \geq c$. Then $\calI$ is $\varepsilon$-fixable. \qed
\end{mytheorem}

The second theorem offers a formal comparison between $\varepsilon$-fixable LexRSM and the state-of-the-art LexRSM variant in the literature, namely GLexRSM~\cite{ChatterjeeGNZZ21}. 
We show the former subsumes the latter. 
In our terminology, GLexRSM is LW-LexRSM that also satisfies the following \emph{expected leftward non-negativity}:
\begin{align*}
                      \forall t \in \bbN. 
        \forall \omega \in \sem{\Lv_t \neq 0}. 
        \forall k \in \br{1, \ldots, \Lv_t(\omega)}.
        \ 
        \bbE[ \mychar_{\sem{k > \Lv_{t+1}}}\cdot\bfX_{t+1}[k] \mid \calF_t](\omega) \geq 0.                             \end{align*}

We note that our result can be also seen as an alternative proof of the soundness of GLexRSM~\cite[Thm. 1]{ChatterjeeGNZZ21}. Our proof is also significantly simpler than the original one, 
as the former utilizes the soundness of UN-LexRSM as a lemma, while the latter does the proof ``from scratch''. 

\begin{mytheorem}[fixable LexRSM generalizes GLexRSM]\label{thm:fixabilityGeneralizesGLexRSM}
    Suppose 
 $\calI$  is a GLexRSM for a stopping time $T$. Then $\calI$ is $\varepsilon$-fixable for any $\varepsilon>0$. \qed
\end{mytheorem} 

Now we move on to a refined variant, $(\varepsilon,\gamma)$-fixability. 
Before its formal definition, we give a theorem that justifies the partial waiving of the ranking condition described in~\S\ref{sect:keyideas}. Below, $\overset{\infty}{\exists}t.\varphi_t$ stands for $\forall k\in \bbN. \exists t\in \bbN.[t > k \land \varphi_t]$.

\begin{mytheorem}[relaxation of the UN-LexRSM condition]\label{thm:relaxedUN-LexRSM}
    Suppose the following are given:
    a probability space $(\Omega, \calF, \bbP)$;
    a filtration $(\calF_t)_{t=0}^\infty$ on $\calF$; and 
    a stopping time $T$ w.r.t. $(\calF_t)_{t=0}^\infty$. 
    Let $\calI=(\seqOf{\bfX_t},\seqOf{\Lv_t})$ be an instance for $T$, and let $\bot \in \bbR$. 
    For each $k \in \br{1,\ldots, n}$, let $\seqOf{\varphi_{t,k}}$ be a sequence of predicates over $\Omega$ such that
\begin{align}
        \overset{\infty}{\exists}t.\varphi_{t,k}(\omega) 
    \Rightarrow 
    \overset{\infty}{\exists}t. [\bfX_t[k](\omega) = \bot \lor k > \Lv_t(\omega)] \quad \mbox{($\bbP$-a.s.)}\label{eq:coinTossArgument}
\end{align}
                    Suppose $\calI$     is an UN-LexRSM with the bottom $\bot$ except that, instead of the ranking condition,  
                $\calI$ satisfies the inequality (\ref{eq:rankingIneq}) 
    only for $t\in \bbN$, $k\in \br{1,\ldots,n}$, and 
        $\omega \in  \sem{ k \leq  \Lv_t \land \lnot(\bfX_t[k]> \bot \land \varphi_{t,k})}$ (with $c=1$).
     Then $T$ is AST w.r.t. $\bbP$. \qed
\end{mytheorem}
The correspondence between the argument in~\S\ref{sect:keyideas} and Thm.~\ref{thm:relaxedUN-LexRSM} is as follows. The predicate $\varphi_{t,k}$ is an abstraction of the situation ``we are at a coin-tossing state at time $t$ in the $k$-th dimension''; and the condition (\ref{eq:coinTossArgument}) corresponds to the infinite coin-tossing argument (for a given $k$, if $\varphi_{t,k}$ is satisfied at infinitely many $t$, then the ranking in the $k$-th dimension is ``done'' infinitely often, with probability 1). 
Given these, Thm.~\ref{thm:relaxedUN-LexRSM} says that the ranking condition of UN-LexRSM can be waived over $\sem{\bfX_t[k]> \bot \land \varphi_{t,k}}$.  
In particular, 
the theorem amounts to the soundness of UN-LexRSM when $\varphi_{t,k} \equiv \mathit{false}$ for each $t$ and $k$.

Based on Theorem~\ref{thm:relaxedUN-LexRSM}, we introduce $(\varepsilon,\gamma)$-fixability as follows. 
There, $\bbP[\varphi\mid \calF'] := \bbE[\mychar_{\sem{\varphi}}\mid \calF']$ is the \emph{conditional probability} of satisfying $\varphi$ given $\calF'$.

\begin{mydefinition}[$(\varepsilon,\gamma)$-fixability]\label{def:egfixability}
    Let $\calI=(\seqOf{\bfX_t},\seqOf{\Lv_t})$ be an instance for $T$, and let $\gamma \in (0,1)$. We call $\calI$ a \emph{$\gamma$-relaxed UN-LexRSM}     for $T$ if $\calI$ satisfies the properties in Thm.~\ref{thm:relaxedUN-LexRSM}, where $\varphi_{t,k}$ is as follows:
    \begin{align}
        \varphi_{t,k}(\omega) \equiv \bbP [\bfX_{t+1}[k]=\bot \mid \calF_t](\omega) \geq \gamma. \label{eq:EGN}
    \end{align}
    We say $\calI$ is \emph{$(\varepsilon,\gamma)$-fixable} if its $\varepsilon$-fixing $\Tilde{\calI}$ is a $\gamma$-relaxed UN-LexRSM.
\end{mydefinition}
The predicate $\varphi_{t,k}(\omega)$ in (\ref{eq:EGN}) is roughly read ``the ranking by $\seqOf{\bfX_t}$ is done at time $t+1$ in dimension $k$ with probability $\gamma$ or higher, given the information about $\omega$ at $t$''. 
This predicate satisfies Condition (\ref{eq:coinTossArgument}); 
hence we have the following corollary, which is the key to the soundness of \emph{lazy LexRSM} in~\S\ref{sect:LLexRSM}. 
\begin{mycorollary}[soundness of $(\varepsilon,\gamma)$-fixable instances]\label{cor:ecfixableWitnessAST}
    Suppose there exists an instance $\calI$ over $(\Omega, \calF, \bbP)$ for a stopping time $T$ that is $(\varepsilon,\gamma)$-fixable for any  $\varepsilon>0$ and $\gamma\in (0,1)$. Then $T$ is AST w.r.t. $\bbP$. \qed
\end{mycorollary}

\section{Lazy LexRSM and Its Soundness}\label{sect:LLexRSM}

Here we introduce another LexRSM variant, \emph{Lazy LexRSM} (LLexRSM). 
We need this variant for our LexRSM synthesis algorithm; 
while $\varepsilon$-fixable LexRSM theoretically answers our key question, 
it is not amenable to LP-based synthesis algorithms because its case distinction makes the resulting constraint nonlinear.

We define LLexRSM map as follows; 
see \emph{Contributions} in~\S\ref{sect:introduction} for its intuitive meaning with an example. 
The definition for an instance is 
\begin{arxiv}
    in Appendix~\ref{append:LLexRSM}. 
\end{arxiv}
\begin{cav}
    in~\cite[Appendix C]{TakisakaZWL24arXiv}.
\end{cav}

\begin{mydefinition}[LLexRSM map]\label{def:LLexRSMmap}
Fix a pCFG $\calC$ with an invariant $I$. 
Let $\boldsymbol{\eta}$ be an MM associated with a level map $\Lv$. 
The MM $\boldsymbol{\eta}$ is called a \emph{Lazy 
LexRSM 
map (LLexRSM map)} over $\calC$ supported by $I$ if 
it is an SC-LexRSM map
over $\calC$ supported by $I$, and satisfies 
\emph{stability at negativity} defined as follows:
\begin{align*}
                &
      \forall\tau \neq \tau_{\mathrm{out}}. 
        \forall s \in \sem{I \land G(\tau)}.
        \forall k \in \br{1,\ldots, \Lv(\tau)-1}. \\
                  &\quad \quad \quad \quad
        \boldsymbol{\eta}[k](s) < 0
        \Rightarrow 
        \forall s' \in \mysucc_\tau(s).  
        \biggl[
            \boldsymbol{\eta}[k](s') < 0 \lor k >\max_{\tau': s'\in \sem{G(\tau')}} \Lv(\tau')
        \biggr].
\end{align*}
\end{mydefinition}

We first observe LLexRSM 
also answers our key question.

\begin{mytheorem}[LLexRSM instantiates SC-LexRF]\label{thm:InstantiationLLexRSM}
Suppose $\boldsymbol{\eta}$ is an SC-LexRSM over a non-probabilistic CFG $\calC$ supported by an invariant $I$, with a level map $\Lv$. 
Then $\boldsymbol{\eta}$ is stable at negativity under $I$ and $\Lv$, and hence, \todo{can be shortened} $\boldsymbol{\eta}$ is an LLexRSM map over $\calC$ supported by $I$, with $\Lv$.  \qed
\end{mytheorem}

Below we give the soundness result of LLexRSM map. We first give the necessary assumptions on pCFGs and MMs, namely \emph{linearity} and \emph{well-behavedness}. 
we say a pCFG is \emph{linear} if the update element of each $\tau \in \Delta_d$ is a linear function (this corresponds to the restriction on PPs to the linear ones); and an MM $\boldsymbol{\eta}$ is \emph{linear} if $\lambda \boldsymbol{x}.\boldsymbol{\eta}(\ell,\boldsymbol{x})$ is linear for each $\ell \in L$. 
We say a pCFG is \emph{well-behaved} if its variable samplings are done via \emph{well-behaved distributions}, which roughly means that their tail probabilities vanish to zero toward infinity quickly enough. 
\begin{arxiv}
    We give its formal definition in the appendix  (Def.~\ref{def:wellBehaved}), 
\end{arxiv}
\begin{cav}
    Its formal definition is given in~\cite[Def. C.4]{TakisakaZWL24arXiv},
\end{cav}
 which is somewhat complex;  
an important fact from the application perspective is that the class of such distributions 
covers all distributions with bounded supports \emph{and} 
some distributions with unbounded supports such as \begin{arxiv}
    the normal distributions 
    (Prop.~\ref{prop:wellBehavednessOfSpecificDist}). 
\end{arxiv}
\begin{cav}
    the normal distributions~\cite[Prop. C.6]{TakisakaZWL24arXiv}.
\end{cav}
Possibly negative (Lex)RSM typically requires some restriction on variable samplings of pCFG (e.g., the \emph{integrability} in~\cite{ChatterjeeGNZZ21})
so that the pre-expectation is well-defined.

The crucial part of the soundness proof is the following theorem, where $(\varepsilon,\gamma)$-fixability takes the key role. Its full proof is given 
\begin{arxiv}
    in Appendix~\ref{append:LLexRSM}. 
\end{arxiv}
\begin{cav}
    in~\cite[Appendix C]{TakisakaZWL24arXiv}.
\end{cav}

\begin{mytheorem}\label{thm:fixabilityOfLLexRSM}
Let $\boldsymbol{\eta}:\calS\to \bbR^n$ be a linear LLexRSM map for a linear, well-behaved pCFG $\calC$. 
Then for any $\Delta$-deterministic scheduler $\sigma$ and initial state $s_I$ of $\calC$, the induced instance is $(\varepsilon,\gamma)$-fixable for some $\varepsilon >0$ and $\gamma \in (0,1)$.
\end{mytheorem}

{\it Proof (sketch).} 
We can show that the $\varepsilon$-fixing $\tilde{\calI} = (\seqOf{\tilde{\bfX}_t},\seqOf{\Lv_t})$ of an induced instance ${\calI} = (\seqOf{{\bfX}_t},\seqOf{\Lv_t})$
almost-surely satisfies 
the inequality (\ref{eq:rankingIneq}) of the ranking condition for each 
$t$, $\omega$, and $k$ 
such that $\tilde{\bfX}_t[k](\omega)=-\varepsilon$ and $1 \leq k \leq \Lv_t(\omega)$~\cite[Prop. C.2]{TakisakaZWL24arXiv}.
Thus it suffices to show, for each $\omega$, $k$, and $t$ such that $\tilde{\bfX}_t[k](\omega)\geq 0$ and $1 \leq k \leq \Lv_t(\omega)$, either 
$\tilde{\calI}$ satisfies the inequality (\ref{eq:rankingIneq}) 
or (\ref{eq:rankingIneq}) as a requirement on $\tilde{\calI}$ is waived 
due to the $\gamma$-relaxation.

Now take any such $t,\omega$, and $k$, and suppose the run $\omega$ reads the program line {\it prog} at time $t$. Then we can show the desired property by a case distinction over {\it prog} as follows. Here, recall $\omega$ is a sequence $s_0s_1\ldots s_ts_{t+1}\ldots$ of program states;
we defined ${\bfX}_t$ by ${\bfX}_t[k](\omega) = \boldsymbol{\eta}[k](s_t)$; and $\bbE[{\bfX}_{t+1}[k] \mid \calF_t] (\omega)$ is the expectation of $\boldsymbol{\eta}[k](s')$, where $s'$ is the successor state of $s_0\ldots s_t$ under $\sigma$ (which is not necessarily $s_{t+1}$).
Also observe the requirement (\ref{eq:rankingIneq}) on $\tilde\calI$ is waived for given $t,\omega$, and $k$ when the value of $\boldsymbol{\eta}[k](s')$ is negative with the probability $\gamma$ or higher.

\begin{enumerate}
    \item Suppose {\it prog} is a non-probabilistic program line, e.g., `$x_i := f(\boldsymbol{x})$' or \mbox{`\textbf{while}} $\varphi$ \mbox{\textbf{do}'}.     Then the successor state $s'$ of $s_t$ is unique.
                If $\boldsymbol{\eta}[k](s')$ is non-negative, then we have 
    $\bbE[\tilde{\bfX}_{t+1}[k] \mid \calF_t] (\omega) = \bbE[{\bfX}_{t+1}[k] \mid \calF_t] (\omega)$,
        so 
    the inequality (\ref{eq:rankingIneq}) is inherited from $\calI$ to $\tilde{\calI}$; 
        if 
        negative, then 
        the requirement (\ref{eq:rankingIneq}) 
    on
    $\Tilde{\calI}$ is waived. The same argument applies to $\mbox{`\textbf{if} $\star$ \textbf{then}'}$  (recall ${\calI}$ is induced from a $\Delta$-deterministic scheduler).
        \item Suppose $\mbox{\it prog} \equiv \mbox{`\textbf{if} \textbf{prob$(p)$} \textbf{then}'}$. 
    By letting $\gamma$ strictly smaller than $p$, we see either $\boldsymbol{\eta}[k](s')$ is never negative,     or it is negative with a probability more than $\gamma$. 
    Thus we have the desired property for a similar reason to Case 1
        (we note this argument requires $p$ to be a constant).
        \item Suppose $\mbox{\it prog} \equiv ` x_i := \mbox{\bf sample}(d)$'. We can show the desired property by taking a sufficiently small $\gamma$; roughly speaking, 
        the requirement (\ref{eq:rankingIneq}) on $\tilde{\calI}$ 
    is waived unless the chance of $\boldsymbol{\eta}[k](s')$ being negative     is very small, in which case the room for ``ill'' exploitation is so small that 
         the inequality (\ref{eq:rankingIneq}) is inherited from $\calI$ to $\tilde{\calI}$. 
            Almost the same argument applies to `$x_i :=  \mbox{\bf ndet}(D)$'.
\end{enumerate}

We note, by the finiteness of program locations $L$ and transitions $\Delta$, we can take $\gamma \in (0,1)$ that satisfies all requirements above simultaneously. \qed

\bigskip

Now we have soundness of LLexRSM as the following theorem, which is almost an immediate consequence of Thm.~\ref{thm:fixabilityOfLLexRSM} and Cor.~\ref{cor:ecfixableWitnessAST}. 

\begin{mytheorem}[soundness of linear LLexRSM map over linear, well-behaved pCFG]\label{thm:FairnessImpliesASTofMaps}
Let $\calC$ be a linear, well-behaved pCFG, and suppose there is a linear LLexRSM map over $\calC$ (supported by any invariant). Then $\calC$ is AST. \qed \end{mytheorem}

\section{Automated Synthesis Algorithm of LexRSM}\label{sect:algorithm}
In this section, 
we introduce a synthesis algorithm of LLexRSM for automated AST verification of linear PPs.
It synthesizes a linear MM in a certain subclass of LLexRSMs. 
We first define the subclass, and then introduce our algorithm. 

Our algorithm is a variant of \emph{linear template-based synthesis}. 
There, we fix a linear MM $\boldsymbol{\eta}$ with unknown coefficients (i.e., \emph{the linear template}), 
and consider an assertion ``$\boldsymbol{\eta}$ is a certificate of AST''; 
for example, in the standard 1-dimensional RSM synthesis, the assertion is ``$\eta$ is an RSM map''. 
We then reduce this assertion into a set of linear constraints via \emph{Farkas' Lemma}~\cite{schrijver1998theory}. 
These constraints constitute an LP problem with an appropriate objective function. A certificate is synthesized, if feasible, by solving this LP problem. 
The reduction is standard, so we omit the details; see e.g.~\cite{TakisakaOUH21}. 

\medskip\noindent{\bf Subclass of LLexRSM for automated synthesis.} 
While LLexRSM resolves the major issue that fixable LexRSM confronts toward its automated synthesis, 
we still need to tweak the notion a bit more, as 
the stability at negativity condition involves the value of an MM $\boldsymbol{\eta}$ in its antecedent part (i.e., it says ``whenever $\boldsymbol{\eta}[k]$ is negative for some $k$...'');  
this makes the reduced constraints via Farkas' Lemma nonlinear. 
Therefore, we augment the condition as follows.

\begin{mydefinition}[MCLC]\label{def:MCLC}
    Let $\boldsymbol{\eta}:\calS\to \bbR^n$ be an MM     supported by an invariant $I$, with a level map $\Lv$. 
    We say $\boldsymbol{\eta}$ satisfies the \emph{multiple-choice leftward condition} (MCLC) if, for each $k\in \br{1,\ldots,n}$, it satisfies either (\ref{item:nonnegative}) or (\ref{item:strongDecrease}) below:     \begin{align}
         & \forall\tau \in \sem{k <\Lv}. 
        \forall s \in \sem{I \land G(\tau)}. 
                 &&  \quad 
         \boldsymbol{\eta}[k](s) \geq 0 ,\label{item:nonnegative} \\
                                           & \forall\tau \in \sem{k <\Lv}. 
        \forall s \in \sem{I \land G(\tau)}. 
        \forall s'\in \mysucc_\tau(s). 
        && \quad
        \boldsymbol{\eta}[k](s') \leq \boldsymbol{\eta}[k](s).\label{item:strongDecrease} 
    \end{align}
                            \end{mydefinition}

Condition (\ref{item:nonnegative}) is nothing but the non-negativity condition in dimension $k$.  
Condition 
(\ref{item:strongDecrease}) augments the ranking condition in the strict leftward of the ranking dimension (a.k.a. the \emph{unaffecting} condition) 
so that the value of 
$\boldsymbol{\eta}[k]$ is non-increasing in the worst-case. 
MCLC implies stability at negativity; 
hence, by Thm.~\ref{thm:FairnessImpliesASTofMaps}, linear SC-LexRSM maps with MCLC certify AST of linear, well-behaved pCFGs. 
They also instantiate SC-LexRFs as follows. 
\begin{mytheorem}[SC-LexRSM maps with MCLC instantiate SC-LexRFs]\label{thm:InstantiationSC-LexRSM+MCLC}
Suppose $\boldsymbol{\eta}$ is an SC-LexRSM map over a non-probabilistic 
CFG $\calC$ supported by $I$, with $\Lv$. 
Then $\boldsymbol{\eta}$ satisfies MCLC under $I$ and $\Lv$. \qed
\end{mytheorem}

\noindent{\bf The algorithm.} 
Our LexRSM synthesis algorithm mostly resembles the existing ones~\cite{ChatterjeeGNZZ21,AgrawalCP18}, so we are brief here; a line-to-line explanation with a pseudocode is 
\begin{arxiv}
    in Appendix~\ref{append:algorithm}. 
\end{arxiv}
\begin{cav}
    in~\cite[Appendix D]{TakisakaZWL24arXiv}.
\end{cav}
The algorithm receives a pCFG $\calC$ and an invariant $I$, and attempts to construct a SC-LexRSM with MCLC over $\calC$ supported by $I$. 
The construction is iterative; 
at the $k$-th iteration, the algorithm attempts to construct a one-dimensional MM $\eta_k$ that ranks transitions of $\calC$ that are not ranked by the current construction $\boldsymbol{\eta} = (\eta_1, \ldots, \eta_{k-1})$, while respecting MCLC. If the algorithm finds $\eta_k$ that ranks at least one new transition, then it appends $\eta_k$ to $\boldsymbol{\eta}$ and goes to the next iteration; otherwise, it reports a failure. Once $\boldsymbol{\eta}$ ranks all transitions, the algorithm reports a success, returning $\boldsymbol{\eta}$ as an AST certificate of $\calC$.

Our algorithm attempts to construct $\eta_k$ in two ways, by adopting either (\ref{item:nonnegative}) or (\ref{item:strongDecrease}) as the leftward condition  at the dimension $k$.
The attempt with the condition (\ref{item:nonnegative}) is done in the same manner as existing algorithms~\cite{ChatterjeeGNZZ21,AgrawalCP18}; 
we require $\eta_k$ to rank the unranked transitions \emph{as many as possible}. 
The attempt with the condition (\ref{item:strongDecrease}) 
is slightly nontrivial; the algorithm demands a user-defined parameter $\mbox{Class}(U) \subseteq 2^U$ for each $U \subseteq \Delta \setminus \br{\tau_{\mathrm{out}}}$. 
The parameter $\mbox{Class}(U)$ 
specifies which set of transitions the algorithm should try to rank, given the set of current unranked transitions $U$; that is, 
for each $\calT \in \mbox{Class}(U)$, the algorithm attempts to find $\eta_k$ that \emph{exactly} ranks transitions in $\calT$.

There are two canonical choices of $\mbox{Class}(U)$.  One is $2^U \setminus \br{\emptyset}$, the brute-force trial; 
the resulting algorithm does not terminate in polynomial time, 
but ranks the maximal number of transitions (by trying each $\calT$ in the descending order w.r.t. $|\calT|$). This property makes the algorithm complete. 
Another choice is the singletons of $U$, i.e., $\br{\br{\tau} \mid \tau \in U}$; 
while the resulting algorithm terminates in polynomial time, it lacks the maximality property. 
It is our future work to verify if there is a polynomial complete instance of our proposed algorithm. Still, any instance of it is complete over yet another class of LLexRSMs, namely linear LW-LexRSMs. 
\begin{arxiv}
The formal statement (Thm.~\ref{thm:SoundnessCompletenessOfAlg}) with proof is 
    in Appendix~\ref{append:algorithm}. 
\end{arxiv}
\begin{cav}
            For a formal statement and its proof, see~\cite[Thm. D.1]{TakisakaZWL24arXiv}. 
\end{cav}

\section{Experiments}\label{sect:experiment}
We performed experiments to evaluate the performance of our proposed algorithm. The implementation is publicly available\footnote{\url{https://doi.org/10.5281/zenodo.10937558}}.

Our evaluation criteria are twofold: 
one is how the relaxed non-negativity condition of our LexRSM---SC non-negativity and MCLC---improves the applicability of the algorithm, 
compared to other existing non-negativity conditions. To this end, we consider two baseline algorithms. \begin{enumerate}[(a)]
\item The algorithm \emph{STR}: 
This is the one proposed in~\cite{AgrawalCP18}, which synthesizes an 
ST-LexRSM.  
We use the implementation provided by the authors~\cite{artifactgit}. \item The algorithm \emph{LWN}: This synthesizes an LW-LexRSM. 
LWN is realized as an instance of our algorithm with $\mbox{\rm Class}(U) = \emptyset$. 
We use LWN as a proxy of the synthesis algorithm of \emph{GLexRSM}~\cite[Alg. 2]{ChatterjeeGNZZ21arxiv}, whose implementation does not seem to exist. 
We note~\cite[Alg. 2]{ChatterjeeGNZZ21arxiv} synthesizes an LW-LexRSM with some additional conditions; 
therefore, it is no less restrictive than LWN.
\end{enumerate}
Another criterion is how the choice of $\mbox{\rm Class}(U)$ affects the performance of our algorithm.  To this end, we consider two instances of it:  \begin{inparaenum}[(a)]
\item \emph{Singleton Multiple Choice} (SMC), given by $\mbox{\rm Class}(U) = \br{\br{\tau} \mid \tau \in U}$; and 
\item \emph{Exhaustive Multiple Choice} (EMC), given by $\mbox{\rm Class}(U) = 2^U \setminus\emptyset$. 
\end{inparaenum}
SMC runs in PTIME, but we do not know if it is complete; EMC does not run in PTIME, but is complete.

We use benchmarks from~\cite{AgrawalCP18},  
which consist of non-probabilistic programs collected in~\cite{alias2010multi} and their probabilistic modifications. The modification is done in two different ways: 
\begin{inparaenum}[(a)]
    \item while loops ``$\textbf{while } \varphi \textbf{ do } P \textbf{ od}$'' are replaced with probabilistic ones ``{\bf while} $\varphi$ {\bf do} ({\bf if prob}$(0.5)$ {\bf then} $P$ {\bf else skip fi) od}''; \label{item:probLoopMod}
    \item in addition to (\ref{item:probLoopMod}), variable assignments ``$x := f(\boldsymbol{x}) + a$'' are replaced with ``$x := f(\boldsymbol{x}) + \mathit{Unif}[a-1, a+1]$''. 
\end{inparaenum}
We include non-probabilistic programs in our benchmark set because the ``problematic program structure'' that hinders automated LexRSM synthesis already exists in non-probabilistic programs (cf. our explanation to  Fig.~\ref{fig:leadingEx}). We also tried two PPs from~\cite[Fig. 1]{ChatterjeeGNZZ21}, which we call {\it counterexStr1} and {\it counterexStr2}. 

We implemented our algorithm  
upon~\cite{AgrawalCP18}, which is available at~\cite{artifactgit}. 
Similar to~\cite{AgrawalCP18}, our implementation 
works as follows: 
\begin{inparaenum}[(1)]
\item it receives a linear PP as an input, and translates it into a pCFG $\calC$;
\item it generates an invariant for $\calC$;
\item via our algorithm, it synthesizes an SC-LexRSM map with MCLC. \end{inparaenum}
Invariants are generated by ASPIC~\cite{feautrier2010accelerated}, and all LP problems are solved by CPLEX~\cite{CPLEX}.

\input{tables/result-tab-129.tex}

\medskip\noindent{\bf Results.} In 135 benchmarks from 55 models, 
STR succeeds in 98 cases, LWN succeeds in 105 cases while SMC and EMC succeed in 119 cases (we did not run STR for {\it counterexStr1} because it involves a sampling from an unbounded support distribution, which is not supported by STR).
Table~\ref{table:instancesWeWin} summarizes the cases where we observe differences in the feasibility of algorithms. As theoretically anticipated, LWN always succeeds in finding a LexRSM whenever STR does; the same relation is observed between SMC vs. LWN and EMC vs. SMC. In most cases, STR, LWN, and SMC return an output within a second\footnote{
There was a single example for which more time was spent, due to a larger size.
}, 
while EMC suffers from an exponential blowup when 
it attempts to rank transitions with Condition (\ref{item:strongDecrease}) in Def.~\ref{def:MCLC}. 
The full results are 
\begin{arxiv}
    in Appendix~\ref{append:fullExperiment}.
\end{arxiv}
\begin{cav}
    in~\cite[Appendix E]{TakisakaZWL24arXiv}.
\end{cav}

On the first evaluation criterion, the advantage of the relaxed non-negativity is evident: 
SMC/EMC have unique successes vs. STR on 21 programs (21/135 = 15.6\% higher success rate) from 16 different models; 
SMC/EMC also have unique successes vs. LWN in 14 programs (14/135 = 10.4\% higher success rate) from 12 models.
This result shows that the program structure we observed in Fig.~\ref{fig:leadingEx} appears in various programs in the real world.

On the second criterion, EMC does not have any unique success compared to SMC. 
This result suggests that SMC can be the first choice 
as a concrete instance of our proposed algorithm. 
Indeed, we suspect that SMC is actually complete---verifying its (in)completeness is a future work.
For some programs, EMC found a LexRSM with a smaller dimension than SMC. 

Interestingly, LWN fails to find a LexRSM for {\it counterexStr2}, despite it being given in~\cite{ChatterjeeGNZZ21} as a PP for which a GLexRSM (and hence, an LW non-negative LexRSM) exists.  
This happens because the implementation in~\cite{artifactgit} translates the PP into a pCFG with a different shape than the one in~\cite{ChatterjeeGNZZ21} (for the latter, a GLexRSM indeed exists); 
the former possesses a similar structure as in Fig.~\ref{fig:leadingEx} 
because different locations are assigned for the {\sf while} loop and {\sf if} branch. 
This demonstrates the advantage of our algorithm from another point of view, i.e., robustness against different translations of PPs.

\section{Related Work}\label{sect:relWork}
There is a rich body of studies in 1-dimensional RSM~\cite{chakarov2013probabilistic,chatterjee2016termination,chatterjee2016algorithmic,chatterjee2017stochastic,ferrer2015probabilistic,mciver2016new,mciver2017new,huang2018new,fu2019termination,moosbrugger2021automated,giesl2019computing},
while lexicographic RSM is relatively new~\cite{AgrawalCP18,ChatterjeeGNZZ21}. 
Our paper generalizes the latest work~\cite{ChatterjeeGNZZ21} on LexRSM as follows:
\begin{inparaenum}[(a)]
\item {\em Soundness of LexRSM as a stochastic process:}\label{item:ComparisonSP} 
soundness of $\varepsilon$-fixable LexRSMs (Def.~\ref{def:epsilonfixing})
generalizes \cite[Thm. 1]{ChatterjeeGNZZ21} in the sense that every GLexRSM is $\varepsilon$-fixable for any $\varepsilon>0$ (Thm.~\ref{thm:fixabilityGeneralizesGLexRSM}); \item {\em Soundness of LexRSM as a function on program states:}\label{item:ComparisonPS} our result (Thm.~\ref{thm:FairnessImpliesASTofMaps}) generalizes \cite[Thm. 2]{ChatterjeeGNZZ21} under the linearity and well-behavedness assumptions; \item {\em Soundness and completeness of LexRSM synthesis algorithms:}\label{item:ComparisonAlg}  our result generalizes the results for one of two algorithms in~\cite{ChatterjeeGNZZ21} that assumes boundedness assumption on assignment distribution~\cite[Thm. 3]{ChatterjeeGNZZ21}.
\end{inparaenum}

\begin{full}We present below a detailed comparison between the latest work~\cite{ChatterjeeGNZZ21} and ours. 
Both \cite{ChatterjeeGNZZ21} and ours present the following theoretical results. 
{\color{blue}\begin{inparaenum}[(a)]
\item {\em Soundness of LexRSM as a stochastic process:}\label{item:ComparisonSP} For this, our result (Thm.~\ref{thm:FCR+SNImpliesAST}) generalizes \cite[Thm. 1]{ChatterjeeGNZZ21}, where we weaken the expected and leftward non-negativity conditions into $\varepsilon$-FCR and SN conditions. 
\item {\em Soundness of LexRSM as a function over program states (i.e., \emph{measurable map}):}\label{item:ComparisonMM} For this, our result (Thm.~\ref{thm:FairnessImpliesASTofMaps}) similarly generalizes~\cite[Thm. 2]{ChatterjeeGNZZ21} in terms of the LexRSM condition;  on the other hand, our theorem requires linearity on LexRSM maps and pCFGs, as well as the boundedness on assignment distributions, while~\cite[Thm. 2]{ChatterjeeGNZZ21} does not. We note that, however, the linearity assumption was necessary for our proof to show a certain fundamental fact (Prop.~\ref{prop:WDAdequacy}.\ref{item:adequacy}) that seems to be also used in the proof of~\cite{ChatterjeeGNZZ21} without a mention.
\item {\em Soundness and completeness of LexRSM synthesis algorithms:}\label{item:ComparisonAlg} In~\cite{ChatterjeeGNZZ21}, soundness is under boundedness assumption on assignment distribution~\cite[Thm. 3]{ChatterjeeGNZZ21}, and completeness is without it~\cite[Thm. 4]{ChatterjeeGNZZ21}; 
our result (Thm.~\ref{thm:SoundnessCompletenessOfAlg}) generalizes the former to the class of LexRSMs with a weaker non-negativity condition.
\end{inparaenum}}
\end{full}

The work~\cite{Huang0CG19} also considers a relaxed non-negativity of RSMs. 
Their \emph{descent supermartingale}, which acts on {\sf while} loops, requires well-foundedness only at every entry into the loop body. 
A major difference from our LexRSM is that they only consider 1-dimensional RSMs; therefore, the problem of relaxing the LW non-negativity does not appear in their setting. 
Compared with their RSM, our LexRSM has an advantage in verifying PPs with a structure shown in Fig.~\ref{fig:leadingEx}, where the value of our LexRSM can be arbitrarily small upon the loop entrance (at some dimension; see $\eta_2$ at $\ell_1$ in Fig.~\ref{fig:leadingEx}).

The work~\cite{mciver2017new} extends the applicability of standard RSM on a different aspect from LexRSM. The main feature of their RSM is that it can verify AST of the symmetric random walk. While our LexRSM cannot verify AST of this process, the RSM by~\cite{mciver2017new} is a 1-dimensional one, which typically struggles on PPs with nested structures. Such a difference can be observed from the experiment result in~\cite{MoosbruggerBKK21} (compare \cite[Table 2]{MoosbruggerBKK21} and {\it nested\_loops}, {\it sequential\_loops} in \cite[Table 1]{MoosbruggerBKK21}).

\section{Conclusion}\label{sect:conclusion}
We proposed 
the first variants of LexRSM that instantiate SC-LexRF.
An algorithm was proposed to synthesize such a LexRSM, and 
experiments have shown that the relaxation of non-negativity contributes applicability of the resulting LexRSM.
We have two open problems: 
one is if the class of well-behaved distributions matches with the one of integrable ones; and  
another is if the SMC variant of our algorithm (see \S\ref{sect:experiment}) is complete. \todo{reconsider}

\section*{Acknowledgment}
We thank anonymous reviewers for their constructive comments on the previous versions of the paper. The term ``ill exploitation'' is taken from one of the reviews that we found very helpful. We also thank Shin-ya Katsumata, Takeshi Tsukada, and Hiroshi Unno for their comments on the paper.

This work is partially supported by National Natural Science Foundation of China No. 62172077 and 62350710215.

\bibliographystyle{splncs04}
\bibliography{ref}

\begin{cav}

%% file: figures/ExForGLexRSM.tex
\lstdefinelanguage{affprob}
{
morekeywords={angel,demon, choice, prob(0.6), prob(0.5), if, then, else, fi, 
while, do, od, 
true, false, and, or, skip, sample},
sensitive = false
}

\lstset{language=affprob}
\lstset{tabsize=1,escapechar=\&}
\newsavebox{\infaa}
\begin{lrbox}{\infaa}
	\begin{lstlisting}[mathescape]
$\ell_1:$
$\ell_2:$
$\quad$
\end{lstlisting}
\end{lrbox}

\lstset{tabsize=2,escapechar=\&}
\newsavebox{\infass}
\begin{lrbox}{\infass}
	\begin{lstlisting}[mathescape]			
skip;
$x:=1$;
...
\end{lstlisting}
\end{lrbox}

\newsavebox{\infastt}
\begin{lrbox}{\infastt}
\begin{lstlisting}[mathescape]
//$\boldsymbol{\eta}=(a_1,\underline{b_1},c_1)$
//$\boldsymbol{\eta}=(\underline{a_2},b_2,c_2)$   
$\quad$
\end{lstlisting}
\end{lrbox}

\begin{figure}[t]%[!htbp]
 %\centering
%\begin{wrapfigure}[6]{r}{0.43\textwidth}
%\vspace{-1em}
	\centering
\scalebox{0.8}{
        \usebox{\infaa}
        %\hspace{0.1cm}
	\usebox{\infass}
	\hspace{0.3cm}
	\usebox{\infastt}
	\hspace{0.1cm}
	% \usebox{\infastinv}
        }
\ 
\scalebox{0.9}{
\small
%%%Long version
% \begin{tabular}{lll}
% \hline
% Non-neg. cond.  & Name of the resulting LexRF   & $\boldsymbol{\eta}$ is non-neg. at \\ \hline
% Strong non-neg. & (Strongly non-neg. LexRF)& $a_1, b_1, c_1, a_2,b_2,c_2$             \\
% LW non-neg.     & Ben-Amram-Genaim (BG-)LexRF    & $a_1, b_1, a_2$                          \\
% SC non-neg.     & Bradley-Manna-Sipma (BMS-)LexRF & $b_1, a_2$                               \\ \hline
% \end{tabular}
%
%%%Short version
\begin{tabular}{rl}
\hline
Non-negativity condition \quad & $\boldsymbol{\eta}$ should be non-neg. at \\ \hline
Strong (ST) non-neg.  \quad   & $a_1, b_1, c_1, a_2,b_2,c_2$             \\
Leftward (LW) non-neg. \quad    & $a_1, b_1, a_2$                          \\
Single-component (SC) non-neg.  \quad   & $b_1, a_2$                               \\ \hline
\end{tabular}
 }
%\vspace{-0.3cm}
    \caption{A demo of different non-negativity conditions for LexRFs. There, the ranking dimensions of the LexRF $\boldsymbol{\eta}$ are indicated by underlines, and the last column of the table shows where each %non-negativity 
    condition requires $\boldsymbol{\eta}$ to be non-negative.} %($b_1$ and $a_2$).} 
\label{fig:ExForGLexRSM}
%\end{wrapfigure}
%\vspace{-0.2cm}
\end{figure}

%% file: figures/LeadingEx.tex
\lstdefinelanguage{affprob}
{
morekeywords={angel,demon, choice, prob(0.6), prob(0.5), if, then, else, fi, 
while, do, od, 
true, false, and, or, skip, sample},
sensitive = false
}

\lstset{language=affprob}
\lstset{tabsize=1,escapechar=\&}
\newsavebox{\infa}
\begin{lrbox}{\infa}
	\begin{lstlisting}[mathescape]

$\ell_1:$
$\ell_2:$
$\ell_3:$

$\ell_4:$

$\ell_5:$
\end{lstlisting}
\end{lrbox}

\lstset{tabsize=2,escapechar=\&}
\newsavebox{\infas}
\begin{lrbox}{\infas}
	\begin{lstlisting}[mathescape]
$x:=0$;
while $x < 5$ do
	if $y < 10$ then
		$y := y + 1$
	else
		$x := x + 1$
	fi
od
$\quad$
\end{lstlisting}
\end{lrbox}

\newsavebox{\infasp}
\begin{lrbox}{\infasp}
	\begin{lstlisting}[mathescape]
$x:=0$;
while $x < 5$ do
	if $y < 10$ then
		$y := y + \mathit{Unif}[1,2]$
	else
		$x := x + \mathit{Unif}[1,2]$
	fi
od
$\quad$
\end{lstlisting}
\end{lrbox}

\newsavebox{\infast}
\begin{lrbox}{\infast}
\begin{lstlisting}[mathescape]

$\boldsymbol{\eta} = (15-2x,  12-y,  1)$
$\boldsymbol{\eta} = (15-2x,  12-y,  0)$             
$\boldsymbol{\eta} = (15-2x,  11-y,     2)$

$\boldsymbol{\eta} = (14-2x, \;\;\;\,\;\;\,\;\; 0,       1)$

$\boldsymbol{\eta} = (\;\;\;\;\;\;\;\;\;\;0, \;\;\,\;\;\,\;\;\; 0,       0)$	
\end{lstlisting}
\end{lrbox}

\newsavebox{\infastinv}
\begin{lrbox}{\infastinv}
	\begin{lstlisting}[mathescape]

$[x<7]$
$[x<5]$
$[y<10, x<5]$

$[y\geq10, x<5]$

$[x\geq 5]$	
	\end{lstlisting}
\end{lrbox}

\begin{figure}[t]%[!htbp]
 \centering
 %\vspace{-0.5cm}
\scalebox{0.8}{
 %\subfloat[\label{fig:leadingEx1}]{
        \usebox{\infa}
        %\hspace{0.1cm}
	%\usebox{\infas}
        \usebox{\infasp}
	\hspace{0.6cm}
	\usebox{\infast}
	\hspace{0.6cm}
	\usebox{\infastinv}
 %}
%\qquad
}
% \scalebox{0.9}{
%  \subfloat[\label{fig:leadingEx2}]{
%         \usebox{\infa}
%         %\hspace{0.1cm}
% 	\usebox{\infasp}
%  }
% }
%
% \subfloat[]{
%   \includegraphics[width=0.1\columnwidth]{figures/InconsisIllust2.pdf}\label{Ex-right}
%  }
\caption{A probabilistic modification of {\sf speedDis1}~\cite{alias2010multi}, where $\mathit{Unif}[a,b]$ is a uniform sampling from the (continuous) interval 
$[a,b]$. %Vectors in the center are the values of the LexRF $\boldsymbol{\eta}$; %$(\eta_1,\eta_2,\eta_3)$; 
%The square brackets 
Inequalities 
on the right represent invariants. While  $\boldsymbol{\eta}$ is not a GLexRSM, it is an LLexRSM we propose; thus it witnesses almost-sure termination of the program.
% (b) is a probabilistic modification of (a), where $\mathit{Unif}[a,b]$ is a uniform sampling from %the interval 
% $[a,b]$.
} 
%\vspace{-0.5cm}
\label{fig:leadingEx}
%\vspace{-0.2cm}
\end{figure}

%% file: figures/CounterEx1.tex
\lstset{language=affprob}
\lstset{tabsize=1,escapechar=\&}
\lstdefinelanguage{affprob}
{
morekeywords={angel,demon, choice, prob(0.6), prob(0.5), if, then, else, fi, 
while, do, od, 
true, false, and, or, skip, sample},
sensitive = false
}

\newsavebox{\infaaa}
\begin{lrbox}{\infaaa}
	\begin{lstlisting}[mathescape]
 
$\ell_1:$
$\ell_2:$
$\ell_3:$
$\ell_4:$

$\ell_5:$
\end{lstlisting}
\end{lrbox}

\lstset{tabsize=2,escapechar=\&}
\newsavebox{\infasss}
\begin{lrbox}{\infasss}
	\begin{lstlisting}[mathescape]	
$x:=0$;$t:=1$;
while $x=0$ do
  $t := t+1$;
  if $\mathbf{prob}(2^{-t})$
    then $x := 1$
  fi
od
$\quad$
\end{lstlisting}
\end{lrbox}

%Later fix the space \,\;\;\;
\newsavebox{\infasttt}
\begin{lrbox}{\infasttt}
\begin{lstlisting}[mathescape]

$\boldsymbol{\eta}=(2-x,\;\;\;\;0,2)$
$\boldsymbol{\eta}=(\;\;\;\;\;\;\,2,\;\;\;\;0,1)$    
$\boldsymbol{\eta}=(\;\;\;\;\;\;\,2,\;\;\;\;0,0)$
$\boldsymbol{\eta}=(\;\;\;\;\;\;\,2,-2^{t},0)$

$\boldsymbol{\eta}=(\;\;\;\;\;\;\,0,\;\;\;\;0,0)$
\end{lstlisting}
\end{lrbox}

 \begin{wrapfigure}[10]{r}{0.45\textwidth}
%\vspace{-1em}
\centering
\scalebox{0.8}{
 % \subfloat[\label{fig:CounterEx1}]{
        \usebox{\infaaa}
        %\hspace{0.1cm}
	\usebox{\infasss}
	\hspace{0.1cm}
	\usebox{\infasttt}
 }
% \quad \
% }
% \scalebox{0.8}{
%  \subfloat[\label{fig:CounterEx2}]{
%         \usebox{\infaaaa}
%         %\hspace{0.1cm}
% 	\usebox{\infassss}
% 	\hspace{0.1cm}
% 	\usebox{\infastttt}
%  }
% }
% \subfloat[]{
%   \includegraphics[width=0.1\columnwidth]{figures/InconsisIllust2.pdf}\label{Ex-right}
%  }
\caption{An example of ``ill'' exploitation. 
} 
\label{fig:CounterEx1}
\end{wrapfigure}
%\end{figure}

%% file: figures/CounterEx2.tex
\lstset{language=affprob}
\lstset{tabsize=1,escapechar=\&}
\lstdefinelanguage{affprob}
{
morekeywords={angel,demon, choice, prob(0.6), prob(0.5), if, then, else, fi, 
while, do, od, 
true, false, and, or, skip, sample},
sensitive = false
}

%%%%Second fig%%%%
\newsavebox{\infaaaa}
\begin{lrbox}{\infaaaa}
	\begin{lstlisting}[mathescape]
 
$\ell_1:$
$\ell_2:$
$\ell_3:$
$\ell_4:$

$\ell_5:$
\end{lstlisting}
\end{lrbox}

\lstset{tabsize=2,escapechar=\&}
\newsavebox{\infassss}
\begin{lrbox}{\infassss}
	\begin{lstlisting}[mathescape]	
$x:=0$;$t:=1$;
while $x=0$ do
  if $\mathbf{prob}(0.5)$
    then $t := 4t$
  else $x := 1$
  fi
od
$\quad$
\end{lstlisting}
\end{lrbox}

%Later fix the space \,\;\;\;
\newsavebox{\infastttt}
\begin{lrbox}{\infastttt}
\begin{lstlisting}[mathescape]

$\boldsymbol{\eta}=(2-x,\;\;\;\;\;t+1)$
$\boldsymbol{\eta}=(\;\;\;\;\;\;\,2,\;\;\;\;\;\;\;\;\;\;\;t)$    
$\boldsymbol{\eta}=(\;\;\;\;\;\;\,
2,
\;\;\;4t+2)$
$\boldsymbol{\eta}=(\;\;\;\;\;\;\,2,-2t-4)$

$\boldsymbol{\eta}=(\;\;\;\;\;\;\,0,\;\;\;\;\;\;\;\;\;\;\;0)$
\end{lstlisting}
\end{lrbox}

 \begin{wrapfigure}[11]{r}{0.45\textwidth}
%\vspace{-1em}
\centering
\scalebox{0.8}{
 % \subfloat[\label{fig:CounterEx2}]{
        \usebox{\infaaaa}
        %\hspace{0.1cm}
	\usebox{\infassss}
	\hspace{0.1cm}
	\usebox{\infastttt}
% }
}
% \subfloat[]{
%   \includegraphics[width=0.1\columnwidth]{figures/InconsisIllust2.pdf}\label{Ex-right}
%  }
\caption{An example of ``harmless'' unbounded negativity. 
} 
\label{fig:CounterEx2}
\end{wrapfigure}
%\end{figure}

%% file: tables/result-tab-129.tex
\begin{table}[t]
\small
\centering
%\vspace{-0.3cm}
\scalebox{0.85}{
\begin{tabular}{|lll|llll|lllllll|}
\hline
\multicolumn{3}{|l|}{\multirow{2}{*}{Benchmark spec.}}                   & \multicolumn{4}{c|}{Synthesis result}                                                         & \multicolumn{3}{l|}{\multirow{2}{*}{Benchmark spec.}}                                        & \multicolumn{4}{c|}{Synthesis result}                                                          \\ \cline{4-7} \cline{11-14} 
\multicolumn{3}{|c|}{}                                                   & \multicolumn{2}{c|}{Baselines}                          & \multicolumn{2}{c|}{Our algs.} & \multicolumn{3}{c|}{}                                                                        & \multicolumn{2}{c|}{Baselines}                           & \multicolumn{2}{c|}{Our algs.} \\ \hline
\multicolumn{1}{|l|}{Model}         & \multicolumn{1}{c|}{p.l.}   & p.a.   & \multicolumn{1}{c|}{STR}   & \multicolumn{1}{c|}{LWN}   & \multicolumn{1}{c|}{SMC}    & EMC   & \multicolumn{1}{l|}{Model}         & \multicolumn{1}{c|}{p.l.}   & \multicolumn{1}{c|}{p.a.}   & \multicolumn{1}{c|}{STR}    & \multicolumn{1}{c|}{LWN}   & \multicolumn{1}{c|}{SMC}    & EMC   \\ \hline
\multicolumn{1}{|l|}{complex}       & \multicolumn{1}{c|}{-} &\multicolumn{1}{c|}{-} & \multicolumn{1}{c|}{$\times$} & \multicolumn{1}{c|}{$\times$} & \multicolumn{1}{c|}{7}      &\multicolumn{1}{c|}{5}     & \multicolumn{1}{l|}{serpent}       & \multicolumn{1}{c|}{-} & \multicolumn{1}{c|}{-} & \multicolumn{1}{c|}{$\times$}  & \multicolumn{1}{c|}{$\times$} & \multicolumn{1}{c|}{3}      &\multicolumn{1}{c|}{3}     \\ \hline
\multicolumn{1}{|l|}{complex}       & \multicolumn{1}{c|}{$\surd$}  &\multicolumn{1}{c|}{-} & \multicolumn{1}{c|}{$\times$} & \multicolumn{1}{c|}{$\times$} & \multicolumn{1}{c|}{7}      &\multicolumn{1}{c|}{5}     & \multicolumn{1}{l|}{speedDis1}     & \multicolumn{1}{c|}{-} & \multicolumn{1}{c|}{-} & \multicolumn{1}{c|}{$\times$}  & \multicolumn{1}{c|}{$\times$} & \multicolumn{1}{c|}{4}      &\multicolumn{1}{c|}{4}     \\ \hline
\multicolumn{1}{|l|}{complex}       & \multicolumn{1}{c|}{$\surd$}  &\multicolumn{1}{c|}{$\surd$}  & \multicolumn{1}{c|}{$\times$} & \multicolumn{1}{c|}{$\times$} & \multicolumn{1}{c|}{3}      &\multicolumn{1}{c|}{3}     & \multicolumn{1}{l|}{speedDis2}     & \multicolumn{1}{c|}{-} & \multicolumn{1}{c|}{-} & \multicolumn{1}{c|}{$\times$}  & \multicolumn{1}{c|}{$\times$} & \multicolumn{1}{c|}{4}      &\multicolumn{1}{c|}{4}     \\ \hline
\multicolumn{1}{|l|}{cousot9}       & \multicolumn{1}{c|}{-} &\multicolumn{1}{c|}{-} & \multicolumn{1}{c|}{$\times$} & \multicolumn{1}{c|}{3}     & \multicolumn{1}{c|}{3}      &\multicolumn{1}{c|}{3}     & \multicolumn{1}{l|}{spdSimMul}     & \multicolumn{1}{c|}{-} & \multicolumn{1}{c|}{-} & \multicolumn{1}{c|}{$\times$}  & \multicolumn{1}{c|}{$\times$} & \multicolumn{1}{c|}{4}      &\multicolumn{1}{c|}{4}     \\ \hline
\multicolumn{1}{|l|}{cousot9}       & \multicolumn{1}{c|}{$\surd$}  &\multicolumn{1}{c|}{-} & \multicolumn{1}{c|}{$\times$} & \multicolumn{1}{c|}{$\times$} & \multicolumn{1}{c|}{4}      &\multicolumn{1}{c|}{4}     & \multicolumn{1}{l|}{spdSimMulDep}  & \multicolumn{1}{c|}{-} & \multicolumn{1}{c|}{-} & \multicolumn{1}{c|}{$\times$}  & \multicolumn{1}{c|}{$\times$} & \multicolumn{1}{c|}{4}      &\multicolumn{1}{c|}{4}     \\ \hline
\multicolumn{1}{|l|}{loops}         & \multicolumn{1}{c|}{-} &\multicolumn{1}{c|}{-} & \multicolumn{1}{c|}{$\times$} & \multicolumn{1}{c|}{$\times$} & \multicolumn{1}{c|}{4}      &\multicolumn{1}{c|}{3}     & \multicolumn{1}{l|}{spdSglSgl2}    & \multicolumn{1}{c|}{$\surd$}  & \multicolumn{1}{c|}{$\surd$}  & \multicolumn{1}{c|}{$\times$}  & \multicolumn{1}{c|}{$\times$} & \multicolumn{1}{c|}{5}      &\multicolumn{1}{c|}{5}     \\ \hline
\multicolumn{1}{|l|}{nestedLoop}    & \multicolumn{1}{c|}{$\surd$}  &\multicolumn{1}{c|}{$\surd$}  & \multicolumn{1}{c|}{$\times$} & \multicolumn{1}{c|}{$\times$} & \multicolumn{1}{c|}{4}      &\multicolumn{1}{c|}{3}     & \multicolumn{1}{l|}{speedpldi3}    & \multicolumn{1}{c|}{-} & \multicolumn{1}{c|}{-} & \multicolumn{1}{c|}{$\times$}  & \multicolumn{1}{c|}{3}     & \multicolumn{1}{c|}{3}      &\multicolumn{1}{c|}{3}     \\ \hline
\multicolumn{1}{|l|}{realheapsort}  & \multicolumn{1}{c|}{-} &\multicolumn{1}{c|}{-} & \multicolumn{1}{c|}{$\times$} & \multicolumn{1}{c|}{3}     & \multicolumn{1}{c|}{3}      &\multicolumn{1}{c|}{3}     & \multicolumn{1}{l|}{speedpldi3}    & \multicolumn{1}{c|}{$\surd$}  & \multicolumn{1}{c|}{-} & \multicolumn{1}{c|}{$\times$}  & \multicolumn{1}{c|}{$\times$} & \multicolumn{1}{c|}{4}      &\multicolumn{1}{c|}{4}     \\ \hline
\multicolumn{1}{|l|}{RHS\_step1}    & \multicolumn{1}{c|}{-} &\multicolumn{1}{c|}{-} & \multicolumn{1}{c|}{$\times$} & \multicolumn{1}{c|}{3}     & \multicolumn{1}{c|}{3}      &\multicolumn{1}{c|}{3}     & \multicolumn{1}{l|}{counterexStr1} & \multicolumn{1}{c|}{-} & \multicolumn{1}{c|}{$\surd$}  & \multicolumn{1}{c|}{N/A} & \multicolumn{1}{c|}{3}     & \multicolumn{1}{c|}{3}      &\multicolumn{1}{c|}{3}     \\ \hline
\multicolumn{1}{|l|}{RHS\_step1}    & \multicolumn{1}{c|}{$\surd$}  &\multicolumn{1}{c|}{$\surd$}  & \multicolumn{1}{c|}{$\times$} & \multicolumn{1}{c|}{3}     & \multicolumn{1}{c|}{3}      &\multicolumn{1}{c|}{3}     & \multicolumn{1}{l|}{counterexStr2} & \multicolumn{1}{c|}{-} & \multicolumn{1}{c|}{$\surd$}  & \multicolumn{1}{c|}{$\times$}  & \multicolumn{1}{c|}{$\times$} & \multicolumn{1}{c|}{4}      &\multicolumn{1}{c|}{4}     \\ \hline
\multicolumn{1}{|l|}{realshellsort} & \multicolumn{1}{c|}{$\surd$}  &\multicolumn{1}{c|}{$\surd$}  & \multicolumn{1}{c|}{$\times$} & \multicolumn{1}{c|}{2}     & \multicolumn{1}{c|}{2}      &\multicolumn{1}{c|}{2}     &                                    &                            &                            &                             &    &                             &       \\ \hline
\end{tabular}
}
\caption{
The list of benchmarks in which a feasibility difference is observed between baselines and proposed algorithms.
Ticks in ``p.l.'' and ``p.a.'' indicate the benchmark has a probabilistic loop and assignment, respectively. 
Numbers in the result indicate that the algorithm found a LexRSM with that dimension; the crosses indicate failures; ``N/A'' means we did not run the experiment.
% Experiment results on algorithms' feasibility. Ticks indicate probabilistic loop and assignment. The numbers indicate the dimensions of LexRSM; the crosses represent failed synthesis. STR failed in all models in the table.
}\label{table:instancesWeWin}
%\vspace{-0.7cm}
\end{table}